\documentstyle[12pt]{article}
\begin{document}

\begin{center}
{\Large {\bf THEORETICAL STUDY OF THE $NN \rightarrow NN \pi \pi$
REACTION}}
\end{center}

\vspace{0.5cm}

\begin{center}
{\Large {L. Alvarez-Ruso$^1$, E. Oset$^1$ and E. Hern\'andez$^2$, }}
\end{center}

\vspace{0.3cm}

{\small {\it
$^1$ Departamento de F\'{\i}sica Te\'orica and IFIC, Centro Mixto Universidad
de Valencia - CSIC, 46100 Burjassot (Valencia) Spain.

$^2$  Departamento de F\'{\i}sica, Universidad de Salamanca, Spain.}}

\vspace{0.7cm}

\begin{abstract}
{\small{
We have developed a model for the $N N \rightarrow N N \pi \pi$
reaction and evaluated cross sections for the different charged channels.
The low energy part of those channels where the pions can be in an isospin
zero state is dominated by $N^*$ excitation, driven by an isoscalar source
recently found experimentally, followed by the decay $N^* \rightarrow
 N (\pi \pi)^{T=0}_{S-wave}$. At higher energies, and in channels where the
pions are not in T=0, $\Delta$ excitation mechanisms become relevant. A
rough agreement with the experimental data is obtained in most channels.   
Repercussions of the present findings
for the ABC effect and the $p p \rightarrow p p \pi^0$ reaction close to
threshold are also suggested.}}
\end{abstract}



\newpage

\section{Introduction}

Pion production in $NN$ collisions is one of the sources of information
on the $NN$ interaction and about nucleon resonance properties. Particularly
the two pion production channel might be especially enlightening in view
of the interesting information obtained from the study of analogous
reactions with two pions in the final state, the $\pi N \rightarrow
\pi \pi N$ and $\gamma N \rightarrow \pi \pi N$ reactions.

The $\pi N \rightarrow \pi \pi N$ reaction close to threshold has been
a testing ground for chiral symmetry \cite{1} in the $\pi N$ sector,
although as one goes away from threshold the contribution of resonances
becomes important \cite{2}. Even at threshold there is an important
non-vanishing contribution from $N^*$ (1440) excitation followed by
the decay $N^* \rightarrow N (\pi \pi)^{T=0}_{S-wave}$ \cite{2}. With
the formalisation of chiral perturbation theory one can further
exploit the ideas of chiral symmetry and perform calculations including
loops with effective Lagrangians \cite{3}. The Roper resonance still plays
and important role but one can use effective Lagrangians involving resonances 
up to order $O (p^2)$, which generalise the coupling of the $N^*  \rightarrow
N (\pi \pi)^{T=0}_{S-wave}$  transition used in Ref. \cite{2} and introduce
one new degree of freedom into the scheme \cite{3}.

In the $\gamma N \rightarrow \pi \pi N$ reaction, which has also been
the subject of study at threshold as a further test of chiral
symmetry \cite{4,5}, one of the most interesting results is the role
of the $N^* (1520, J^\pi = 3/2^-)$ resonance \cite{6}.
The mechanism of $N^*$ (1520) photo-excitation followed by the decay
into $\Delta \pi$ interferes with the dominant Kroll Ruderman
$\Delta N \pi \gamma$ term and allows one to obtain information
about the $N^*$ (1520) $\rightarrow \Delta \pi$ decay amplitudes, additional
to the one obtained from the analysis of the $\pi N \rightarrow
\pi \pi N$ reaction \cite{7}, and which poses new challenges
to quark models of baryons \cite{8}.
Roper excitation also plays an important role at threshold in the
$\gamma N \rightarrow \pi \pi N$ reaction but its role is shadowed by the
contribution of other resonances at higher energy \cite{9}.

A common feature of the $\pi N \rightarrow \pi \pi N$ and $\gamma N
\rightarrow \pi \pi N$ reactions at threshold is the
role of the $N^* $ (1440)  followed by its decay into
$N (\pi \pi)^{T=0}_{S-wave}$ . This gives us a 
hint that Roper excitation might also play an important role in the
$N N \rightarrow N N \pi \pi$ reaction close to threshold. On the other
hand, contrary to the case of $N^*$ excitation from $\pi N$ and $\gamma
N$, which is well known, now we have to deal with the
$N N \rightarrow N N^*$ transition which is not so well explored as
the one baryon vertices. An important step in understanding  
the $NN \rightarrow
N N ^*$ reaction was given with the experiment of Ref. \cite{10}
which showed a relatively large strength for $N^*$ excitation due to the
exchange of an isoscalar source, the only one allowed in the
($\alpha, \alpha'$) reaction on the proton when this proton
is excited to the Roper. The experiment has been analysed in Ref. 
\cite{11}, where the large background from delta excitation in the 
projectile and an important interference term between the latter mechanism
and the one from Roper excitation in the target are properly accounted
for. The analysis provides the strength for the isoscalar $NN \rightarrow
NN^*$ transition which turns out to be large compared to ordinary one pion
exchange.

The combination of this $N^*$ isoscalar excitation with the $N^*
\rightarrow N (\pi \pi)^{T=0}_{S-wave}$ decay seems then called to play
an important role in the $N N \rightarrow N N \pi \pi$ reaction and
this will be the case, as we shall see. On the other hand in the 
$N N \rightarrow N N \pi \pi$ reaction there are many terms which contribute,
starting from the set of terms provided by chiral Lagrangians, plus
terms mediated by $\Delta$ resonance excitation. We study all 
these terms  asserting their relevance as a function of energy. 
Furthermore we observe that the weight of the different terms varies
appreciably from channel to channel and hence the
combined information from these channels puts  strong  tests to the model.

The comparison with the limited experimental information available on
this reaction \cite{12n,13n,14n} shows that a rough agreement in most channels 
can be obtained. Remaining discrepancies in some channels are discussed.

\section{Contribution from non-resonant terms}

The model that we use is depicted diagrammatically in Fig. 1 and contains
terms proceeding through baryon resonance excitation and other non-resonant
terms. The latter ones are shown in  diagrams (1), (2), (3) and require the
knowledge of the $\pi N \rightarrow \pi N$ amplitude, the $\pi N N$ vertex, 
the $\pi \pi \rightarrow \pi \pi$ amplitude and the $N N \pi \pi \pi$
vertex. In the spirit of the phenomenological approach followed here, we
take the empirical $\pi N \rightarrow \pi N$ amplitude. The other ingredients
are derived from the lowest order SU(2) chiral Lagrangians at tree 
level \cite{12, 13}. 

The lowest chiral Lagrangians which we need are given by

\begin{equation}
{\cal L} = {\cal L}_2 + {\cal L}_1^{(B)} 
\end{equation}

\noindent
where ${\cal L}_2$ is the Lagrangian in the meson sector and 
${\cal L}_1^{(B)}$ contains the baryon sector with the meson baryon
interactions. We have

\begin{equation}
{\cal L}_2 = \frac{f^2}{4} < \partial_\mu U^\dagger \partial^\mu U +
M ( U  + U^\dagger) >
\end{equation}

\begin{equation}
{\cal L}_1^{(B)} = \bar{\Psi} ( i \gamma^\mu \nabla_\mu - M_B + 
\frac{g_A}{2} \gamma^\mu \gamma_5 u_\mu ) \Psi \, ,
\end{equation}

\noindent
where the symbol $< >$ denotes the flavour trace of the $SU (2)$ matrices
and $U, u$ are $2 \times 2$ matrices 

\begin{equation}
U (\phi) = u (\phi)^2 = exp \, ( i \sqrt{2} \Phi /f)
\end{equation}

\noindent
where

\begin{equation}
\begin{array}{ll}
\Phi (x) & \equiv  \frac{1}{\sqrt{2}}\, \vec{\tau}\, \vec{\phi}\\[2ex]
& = \left( \begin{array}{cc}
\frac{1}{\sqrt{2}} \pi^0 & \pi^+\\
\pi^- & - \frac{1}{\sqrt{2}} \pi^0 
\end{array} \right) \end{array}
\end{equation}
and
\begin{equation}
\Psi = \left( \begin{array}{cc} p\\ n \end{array} \right) \, .
\end{equation}
The mass matrix $M$ appearing in Eq. (2), in the limit of $m_u = m_d$, is 
 
\begin{equation}
M = \left( \begin{array}{cc}
m^2_\pi & 0\\
0 & m_\pi^2
\end{array} \right) \, .
\end{equation}
Furthermore, $\nabla_\mu$ is a covariant derivative defined as

\begin{equation}
\begin{array}{l}
\nabla_\mu \Psi = \partial_\mu \Psi + \Gamma_\mu \Psi \, ,\\[.4cm]
\Gamma_\mu = \frac{1}{2} (u^\dagger \partial_\mu u + u \partial_\mu u^\dagger
) \, ,\\[.4cm]
u_\mu =  i u^\dagger \partial_\mu U u^\dagger \, .
\end{array}
\end{equation}
By expanding the Lagrangian ${\cal L}_2$ up to order
$ O (\phi^4)$ and keeping the interaction terms in ${\cal L}_{1}^{(B)}$ up
to order $O (\phi^3)$ as done in Ref. \cite{16}, we obtain the diagrams (1), 
(2) and (3) of  Fig. 1 which contribute to the $N N \rightarrow N N \pi \pi$ 
reaction.

The term with $\Gamma_\mu \Psi$ in ${\cal L}_1^{(B)}$ contributes to
the isovector part of the s-wave $\pi N$ amplitude. The isoscalar part of this
amplitude is zero at this order in the chiral expansion and empirically is
much smaller than the isovector one. The isoscalar amplitude is generated when
terms of order $ O (p^2)$ in the baryon chiral Lagrangian are introduced
\cite{3}. In this latter reference the couplings of the new Lagrangians are 
deduced from empirical information of the $\pi N$ amplitude in Ref. \cite{3}. 
With the same philosophy we take the isoscalar  $\pi N$ amplitude from experiment and add
it to the vector part. It is customary to write the isoscalar $\pi N 
\rightarrow \pi N$ amplitude in terms of $\lambda_1$. Using Mandl and Shaw 
normalisation \cite{17n} ( $-i t \equiv {\cal M}$ of Mandl and Shaw ) one has

\begin{equation}
- i t^{(s)} = - i 4 \pi \frac{2 \lambda_1}{\mu} \quad ; \quad
\lambda_1 = 0.0075 \quad ; \quad \mu \equiv m_\pi
\end{equation}

\noindent
which  is diagonal in spin and isospin. The isovector part for $p \pi^+ 
\rightarrow p \pi^+$ amplitude is given by

\begin{equation}
- i t^{(v)} = - i \frac{1}{4 f^2} \bar{p} \gamma^\mu p ( p_\mu +
p'_\mu )
\end{equation}

\noindent
where $p, p'$ are the initial, final pion four-momenta and $\bar{p}$ ($p$) 
besides the $\gamma^\mu$ matrix stand for the spinor of the final (initial)
protons. This would compare with the usual empirical form of the low energy 
amplitude

\begin{equation}
- i t^{(v)} = - i 4 \pi \frac{\lambda_2}{\mu^2} (p + p')^0
\end{equation}

\noindent
which comes from Eq. (10) by retaining the dominant zero component, with
the equivalence $\lambda_2 = \mu^2 / 16 \pi f^2$, which holds up to
15$\%$. We divert from the standard chiral expansion and use the
empirical $\pi N$ amplitudes in our evaluations. 

Eq. (10) is trivially modified in the other isospin channels: same 
factor in $n \pi^- \rightarrow n \pi^-$, relative minus sign in $\pi^+ n 
\rightarrow \pi^+ n$ and $\pi^- p \rightarrow \pi^- p$, zero contribution
for $\pi^0 p (n) \rightarrow \pi^0 p (n)$, extra factor $\sqrt{2}$
for $\pi^- p \rightarrow \pi^0 n$ and $\pi^0 n \rightarrow \pi^- p$
and extra factor $- \sqrt{2}$ in $\pi^0 p \rightarrow \pi^+ n$ and
$\pi^+ n \rightarrow \pi^0 p$.

The amplitudes $\pi \pi \rightarrow \pi \pi$ come from the expansion
of ${\cal L}_2$ up to order $O (\phi^4)$. In the diagram (1)
of Fig. 1 they appear with two pions off shell. We need four
charged amplitudes from where we deduce the rest by using crossing
symmetry. We shall give the explicit expressions only in 
the $\pi^0 \pi^0 \rightarrow \pi^+ \pi^-$ reaction. The expressions for the
other channels can be obtained in a similar way.
We get

\begin{equation}
- i t = i \frac{1}{3 f^2}  ( 3 m^2_{\pi} + 4 p^+ \cdot p^-
+ 2 q \cdot q' )
\end{equation}

\noindent
where $q, q'$ are the $\pi^0$ momenta and $p^+, p^-$ the momenta of
$\pi^+$ and $\pi^-$ respectively.

The other vertex needed is the $\pi N N $ which comes from the
$g_A$ terms of Eq. (3) expanding up to
$O (\phi )$. For $\pi^0 p p $ and an outgoing $\pi^0$
with momentum $q$ it is given by

\begin{equation}
- i t = \frac{g_A}{2} \frac{1}{f} \, \bar{p} \gamma^\mu \gamma_5 p q_\mu
\end{equation}

\noindent
which in the non-relativistic description is $- \frac{ f_{\pi N N}}{\mu}
\vec{\sigma} \vec{q}$, the standard Yukawa coupling, with the equivalence 
$g_A / 2 f = f_{\pi N N} /\mu$. 

Finally the other ingredient needed is the three pion vertex which comes from
the expansion of the $g_A$ term in Eq. (3) up to
order $O (\phi^3)$. For $\pi^0 p \rightarrow p \pi^+ \pi^-$, with $q'$ the
$\pi^0$ momentum we obtain 

\begin{equation}
- i t = \frac{g_A}{2} \bar{p} \gamma^\mu \gamma_5 p
\frac{1}{12 f^3} (2 p^+_{\mu} + 2 p^-_{\mu} +
4 q'_\mu)
\end{equation}

\noindent
and similar expressions in other isospin channels.

With these ingredients one constructs immediately the contribution
from diagrams (1)-(3) of Fig. 1 for the $p p \rightarrow p p \pi^+
\pi^-$ case by following the standard Feynman rules. As an example,
the term corresponding to the diagram (1) of Fig. 1
for $p p \rightarrow p p \pi^+ \pi^-$ is given by

\begin{equation}
\begin{array}{ll}
- i t = & i \frac{1}{3f^2} (3 m_{\pi}^2 + 4 p^+ \cdot
p^- + 2 q \cdot q' ) \, . \\[2ex]
& \frac{1}{q^2 - m_{\pi}^2} \frac{1}{q'^2 - m_{\pi}^2} (\frac{g_A}{2}
\frac{1}{f})^2 \left( \bar{p} \gamma^\mu \gamma_5 p q_\mu \right)_1 
\left( \bar{p} \gamma^\nu \gamma_5 p q'_\nu \right)_2
\end{array}
\end{equation}

\noindent
where the indices 1 and 2 refer to the first and  second nucleons 
respectively. In the Appendix we further reduce this
expression to a Pauli spinor notation and care explicitly about baryon 
antisymmetry, the symmetry of the diagrams and off shell form
factors. 

\section{ $\Delta$ resonance terms contribution}

Next we introduce terms with $\Delta$ excitation. We need the $\pi N \Delta$
coupling, whose vertex is given in a Pauli spinor notation by

\begin{equation}
- i \delta \tilde{H}_{\pi N \Delta} = \frac{f^*}{\mu} \vec{S}\,^\dagger
\vec{q} \, T^{\dagger \lambda} \quad + \quad h.c.
\end{equation}

\noindent
corresponding to the vertex $\pi^{\lambda} + N \rightarrow \Delta$,
with $\lambda$ the isospin
index of the pion, $q$ its momentum
and $\vec{S}\,^\dagger$ ( $\vec{T}\,^{\dagger}$ ) the
spin ( isospin ) transition  operators from 1/2 to 3/2 with the 
normalisation

\begin{equation}
\left\langle \frac{3}{2}, M \right| S^\dagger_\nu \left|\frac{1}{2}, m 
\right\rangle  
=  C \left( \frac{1}{2}, 1, \frac{3}{2} ; m, \nu, M \right)
\end{equation}

\noindent
where $\nu$ is the spherical component of $ \vec{S}\,^\dagger$ and
same normalisation for $\vec{T}\,^\dagger$.

We shall also need the $\pi \Delta \Delta $ vertex for $\pi \Delta
\rightarrow  \Delta $ given by 

\begin{equation}
- i \delta \tilde{H}_{\pi \Delta \Delta} = \frac{f_\Delta}{\mu}
\vec{S}_\Delta \vec{q} \; T_\Delta^\lambda \quad  
\end{equation}

\noindent
where now $\vec{S}_\Delta$ ( $\vec{T}_\Delta$ )
are the ordinary spin ( isospin ) matrices for spin ( isospin ) 3/2. Two
properties involving sums over intermediate $\Delta$ spins are needed
in the evaluation of the diagrams and they are

\begin{equation}
\sum_{M} S_i \left| M_s \right\rangle \left\langle M \right| S^\dagger_j = 
\frac{2}{3} \delta_{ij} - \frac{i}{3} \epsilon_{ijk} \sigma_k
\end{equation}

\begin{equation}
\sum_{M, M'}  S_i \left| M' \right\rangle 
\left\langle M' \right| S_{\Delta,j} \left| M \right\rangle
\left\langle M \right| S^\dagger_k =  \\[2ex]
 \frac{5}{6} i \epsilon_{ijk} - \frac{1}{6} \delta_{ij} \sigma_k
+ \frac{2}{3} \delta_{ik} \sigma_j - \frac{1}{6} \delta_{jk}
\sigma_i
\end{equation}

For the coupling constants we take, $f^{*2}/4\pi = 0.36, \quad
f_\Delta/f_{\pi NN} = \frac{4}{5}$, where the first one is empirical and
the second one comes from the quark model. The couplings in Eq. (16) (18)  
have to be evaluated in the rest frame of the created $\Delta$.

With these new ingredients and those of the former section one can
evaluate the diagrams (9)-(15) which appear in Fig. 1. The expressions for the
amplitudes of the relevant terms are written in the Appendix. As
we shall see, all these terms are small except  diagram (12), the one
involving the excitation of a $\Delta$ in each nucleon. This diagram
is only relevant at energies $T_p > 1 \; GeV$ in the laboratory frame,
where the terms involving crossed $\Delta$ are small ( diagrams (13)-(15) ).
At lower energies all of them become negligible since they involve
two p-wave couplings of the pions ( diagrams (10) and (11) involve only one
p-wave coupling but they are small anyway since the s-wave coupling is
comparatively small ).

One warning is however in order with respect to diagrams (9), (12)-(15) in 
Fig. 1 which involve  two p-wave couplings. When this is the case one must 
take into account the indirect effect of the rest of the $NN$ interaction,
particularly the repulsive force at short distances. This is so because
the combination of the pion propagator and the p-wave couplings
contains a $\delta(\vec{r})$ force which becomes inoperative in the
presence of short range correlations. We follow a standard procedure
to account for this which is to substitute

\begin{equation}
\hat{q}_i \hat{q}_j D (q) \rightarrow V'_L (q) \, \hat{q}_i \hat{q}_j
+ V'_T (q) (\delta_{ij} - \hat{q}_i \hat{q}_j)
\end{equation}

\noindent
which substitutes the pure spin longitudinal pion exchange by a 
mixture of spin longitudinal and transverse contributions. Following
Ref. \cite{17} we take

\begin{equation}
\begin{array}{c}
V'_L (q) = V'_\pi (q) + g'_l (q)\, ,\\[2ex]
V'_T (q) = V'_\rho (q) + g'_t (q)\, ,\\[2ex]
V'_\pi (q) =F^2_\pi (q^2) \vec{q}\,^2 D_\pi (q) \, , \\[2ex]
V'_\rho (q) = F^2_\rho (q^2) \vec{q}\,^2 D_\rho (q) C_\rho \, ,\\[2ex]
g'_l (q) = - (\vec{q}\,^2 + \frac{1}{3} q^2_c) \tilde{F}\,^2_\pi
\tilde{D}_\pi\,^2 - \frac{2}{3} q^2_c \tilde{F}_\rho\,^2 \tilde{D}_\rho
C_\rho \, ,\\[2ex]
g'_t (q) = -\frac{1}{3} q^2_c \tilde{F}_\pi\,^2 \tilde{D}_\pi - (\vec{q}\,^2
+ \frac{2}{3} q^2_c) \tilde{F}_\rho\,^2 \tilde{D}_\rho C_\rho \, 
\end{array}
\end{equation}

\noindent
where $D_{\pi,\rho}(q)$ are the meson propagators and $F_{\pi,\rho}(q)$ are
meson form factors of the monopole type with $\Lambda_{\pi}= 1.3 \ GeV$,
$\Lambda_{\rho}= 1.4 \ GeV$ and $C_\rho=3.94$, consistent with the Bonn model
\cite{18}. The corresponding functions with a tilde are obtained by
substituting, in the argument $q$ of the function, $\vec{q}\,^2$ by
$\vec{q}\,^2+q_c^2$, where $q_c=780 \ MeV$, the inverse of the short range 
correlations scale.  

\section{ $N^* (1440)$  terms contribution}

Next we introduce the $N^* \, (1440)$ excitation. The $N^*$ couples
to $N \pi\,, \Delta \pi \, , N (\pi \pi)^{T=0}_{S-wave}$  as
important decay channels. The corresponding vertices are given by

\begin{equation}
- i \delta \tilde{H}_{\pi N N^*} = \frac{\tilde{f}}{\mu} \vec{\sigma}
\vec{q} \, \tau^\lambda \, + \, h.c.
\end{equation}

\begin{equation}
- i \delta \tilde{H}_{\pi N^* \Delta} = \frac{g_{\pi N^* \Delta} }{\mu}
\vec{S}\,^\dagger \vec{q} \; T^\lambda
\end{equation}

\noindent
both evaluated in the $N^*$ rest frame. The coupling constants are $\tilde{f}
= 0.477$ and $g_{\pi N^* \Delta} = 2.07$ \cite{8}.

For the $N^* \rightarrow N (\pi \pi)^{T=0}_{S-wave}$ decay channel, which
accounts for $5 - 10 \%$ of the total width according to the particle data
table \cite{30}, we take the $O (p^2)$
Lagrangian as given in Ref. \cite{3} using an SU(2) formalism

\begin{equation}
{\cal L}_{N^* N \pi \pi} = c^*_1
\bar{\psi}_{N^*} \chi_+ \psi_N - \frac{c^*_2}{M^* \,^2 }(\partial_\mu
\partial_\nu \bar{\psi}_{N^*} ) u^\mu u^\nu \psi_N + h. c.
\end{equation}

\noindent
where

\begin{equation}
\chi_+ = \mu^2 (2 - \frac{\vec{\phi}\,^2}{f^2} + \ldots ) \, ; \, u_\mu = 
- \frac{1}{f} \vec{\tau} \partial_\mu \vec{\phi} + \ldots
\end{equation}

\noindent
and $\vec{\phi}$ stands  for the pion field. Hence, expanding up to second 
order in the pion fields and keeping the largest $\mu, \nu = 0, 0$ components,
which neglects terms of order $O (\frac{p^2}{{M^*}^2})$, we find

\begin{equation}
{\cal L}_{N^* N \pi \pi} = - c^*_1 \mu^2 \bar{\psi}_N
\frac{\vec{\phi}\,^2}{f^2} \psi_N + c^*_2 \bar{\psi}_{N^*} 
\frac{1}{f^2} (\vec{\tau} \partial_0 \vec{\phi}) 
(\vec{\tau} \partial_0 \vec{\phi}) \psi_N + h. c.
\end{equation}
This Lagrangian leads to an effective vertex 

\begin{equation}
- i \delta \tilde{H}_{N^* N \pi \pi} = - i\,2\, \frac{\mu^2}{f^2} 
\left( c^*_1 + c^*_2 \frac{\omega_1 \omega_2}{\mu^2} \right)
\end{equation}
for $\pi^+ \pi^-$ and $\pi^0 \pi^0$ creation and zero otherwise. Here
$\omega_1,\omega_2$ are the energies of the pions.  

Following the steps of Ref. \cite{3} we fit the experimental
width from $\pi^0 \pi^0$ plus $\pi^+ \pi^-$ final states

\begin{equation}
\Gamma_{N^* \pi \pi} = \alpha (c^*_1)^2 + \beta (c^*_2)^2 + \gamma c^*_1 c^*_2 
\end{equation}

\noindent
and we take $\Gamma_{N^* N \pi \pi}$ assuming $\Gamma^*_{N^* \, tot} =
350 \, MeV$ \cite{30} and branching ratio of 7.5$\%$ for the $N
(\pi \pi)^{T=0}_{S-wave}$ decay channel. We obtain $\alpha = 0.497\,
10^{-3}\, GeV^3$, $\beta = 3.66\, 10^{-3}\, GeV^3$ and $\gamma = 2.69\, 10^{-3}
\, GeV^3$. These values are somewhat different from those of Ref. \cite{3} 
where a smaller width for the Roper was taken. With the $c^*_1$ parameter 
alone one gets a solution for that parameter, but with the two parameters one 
gets an ellipse in $c^*_1$, $c^*_2$. This ellipse spans over a large range of 
values of $c^*_1$, $c^*_2$. It is clear that one needs further constrains in 
order to determine $c^*_1$ and $c^*_2$. We use the $\pi^- p \rightarrow \pi^+ 
\pi^- n$ reaction in order to impose such constrains. For this purpose we use 
the model of  Ref. \cite{2}. The cross section  shows a strong sensitivity 
to the values of $c^*_1$, $c^*_2$ and spans about two orders of magnitude when 
the values of these parameters are varied along the ellipse of
Eq. (29) ( details can be seen in Ref. \cite{31} ). The best agreement with
the experiment is obtained with $c^*_2 = 0$, which corresponds to 
$c^*_1 = -7.27\, GeV^{-1}$ ( set I ). However, the experimental errors are 
compatible with the use of 
$c^*_1 = -12.7\, GeV^{-1}$, $c^*_2 = 1.98\, GeV^{-1}$  ( set II ) and 
intermediate values in the ellipse. In order to illustrate the uncertainties 
in $c^*_1$, $c^*_2$ we plot the cross sections using the extreme values of 
these parameters, compatible with
the $\pi^- p \rightarrow \pi^+ \pi^- n$ cross sections, quoted above. A best
fit to the different $\pi N \rightarrow \pi \pi N$ channels is done in
\cite{32} using a Lagrangian for the $N^* \rightarrow N (\pi
\pi)^{T=0}_{S-wave}$ process, equivalent to the one of Eq. (27) at low
energies. Although the data would roughly be compatible with the use
of $c^*_2 = 0$, they obtain a best fit with other set of parameters.
However, this work contains some small differences with respect to the model
of Ref. \cite{2}, like the relativistic treatment, the use of
an $N^*$ width assuming phase space for $\pi N$ decay only and the neglect of
the isoscalar $\pi N$ amplitude and intermediate $\Delta \Delta$ states 
( which are included in \cite{2} ). These are
small differences, but they have some influence in the $c^*_1$ and $c^*_2$
parameters. For the reason of consistency with our input, we take the values
of $c^*_1$ and $c^*_2$ from our own analysis.    

The diagrams which we obtain now with these new ingredients are (4)-(8) of
Fig. 1. Diagrams (4) and (5) are relevant but (6) and (7) give a much smaller
contribution. Diagram (8) is also important because both $N^*$ and $\Delta$
can be placed simultaneously on shell. Note that we disregard terms like (8)
or (9) with crossed $N^*$ or $\Delta$ poles.
Also in the case of the $N^*$ we disregard the term with two $N^*$ excitations,
both in the same nucleon or in different ones. Because the $\pi N N$
coupling is about one fourth of the $\pi N \Delta$ one, and the $N^*$
propagator would be placed on shell at higher energies, these terms are
small compared to the corresponding ones with two $\Delta$ or $N^* \Delta$
excitation in the energy range where we are.  

In Fig 1. we have explicitly shown what kind of particle exchange we consider.
For instance, in diagrams (6), (7) we exchange one of the pions coming from
the $N^*$ decay. In diagrams (4), (5) and (8) we consider the effective
isospin T=1 interaction of Eq. (21), but in addition we must consider an
exchange in the T=0 channel, to which we come below.

The analysis of the $(\alpha, \alpha')$ reaction on a proton target 
\cite{10} carried out in Ref. \cite{11} interpreted the results
in base of two mechanisms: $\Delta$ excitation in the projectile depicted
in Fig. 2 (a) and $N^* $ (1440) excitation in the proton target, Fig 2 (b). 
However, an important interference was found between the  mechanism of $\Delta$
excitation in the projectile and $N^*$ excitation in the target 
followed by the decay of the $N^*$ into $N \pi$. Since the $^4 He$ beam
has $T = 0$ the $N^*$ excitation in the $(\alpha, \alpha')$
reaction requires the exchange of an isoscalar object.
In meson exchange pictures it could be $\sigma$ or $\omega$ exchange.
However, the experiment has not enough information to provide the
separate strength of both ingredients and only the
strength of the combined exchange can be extracted. This transition
amplitude was parametrised in Ref. \cite{11} in terms of an
effective ``$\sigma$'' which couples to $N N$ as the $\sigma$ exchange
of the Bonn model \cite{18} and couples to $N N^*$ with an unknown
strength which is  determined by a best fit to the data. Hence, we have

\begin{equation}
\begin{array}{c}
- i \delta \tilde{H}_{\sigma \, N N} = F_\sigma (q) g_{\sigma NN} \quad ;
\quad g^2_{\sigma N N} /4 \pi = 5.69\\[2ex]
- i \delta \tilde{H}_{\sigma \, N N^*} = F_\sigma (q) g_{\sigma N N^*}
\end{array}
\end{equation}

\noindent
where $F_\sigma (q)$ is a form factor of the monopole type with
$\Lambda_\sigma = 1700 \, MeV$, assumed equal in both vertices.
The fit to the data provides a value $g^2_{\sigma N N^*} /4 \pi = 1.33$,
with $g_{\sigma N N}$ and  $g_{\sigma N N^*}$ of the same sign.

In our selection of the relevant diagrams we have excluded those with
a nucleon pole. These terms are smaller than those with $\Delta$ pole
because the $\pi N N$ coupling 
is about a factor two smaller than the $\pi N \Delta$
one, and also because there are usually large cancellations between direct
and crossed nucleon pole terms. Indeed, assume we have the diagrams (12) and
(14) of Fig 1 substituting the left $\Delta$ by a nucleon. For small
momenta of the outgoing pion, on the left of the diagram we would have for
the sum of the two nucleon propagators  

\begin{equation}
G^{(a)} + G^{(b)}  \simeq \frac{1}{M + m_\pi - M} +
\frac{1}{E (\vec{p}\,')  - m_\pi - E (\vec{p}\,)} \simeq 0 \, .
\end{equation}
This cancellation would not occur if we had a resonance in the
intermediate state since there is a difference of masses and now we
would have

\begin{equation}
G^{(a)}_R + G^{(b)}_R \simeq \frac{1}{M + m_\pi - M_R} +
\frac{1}{E (\vec{p}) - m_\pi - E_R (\vec{p})} \simeq 
\frac{1}{m_\pi - \Delta M} + \frac{1}{- m_\pi - \Delta M} \, .
\end{equation}

We have also omitted terms involving the exchange of vector mesons in
diagrams of the type of 
(1)-(3) in Fig. 1. We estimate their contribution to be small as follows.
Take for instance the equivalent of diagram (1) substituting the virtual
pions by virtual $\rho$ mesons. A general expression for the $\rho \rho \pi \pi$
Lagrangian is given in \cite{33}. For our purpose we can get the low energy
estimate by using the analogy of the minimal coupling generation of the
$\gamma \gamma$ coupling to $\pi \pi$, in which case we obtain a Lagrangian
of the type 

\begin{equation}
{\cal L}_{\rho \rho \pi \pi} = 
\frac{f_{\rho}^2}{2} (\vec{\rho}_{\mu} \cdot \vec{\rho}^{\mu}) 
(\vec{\phi} \cdot \vec{\phi})
\end{equation}
with $f_{\rho} = 6.1$, the $\rho \pi \pi$ coupling. The diagram (1) with two
virtual $\rho$ mesons is now readily evaluated and we find at the end that
its strength is about the same as the corresponding diagram with two virtual
pions. Short range correlations also help to further reduce the $\rho \rho$
diagram versus the $\pi \pi$ one. The fact that these meson-meson terms are 
found to be small altogether suggests that this should also be the case with 
the inclusion of the $\rho$ meson in the analogous diagrams.

As for the $\omega$ exchange, one can think of a diagram like (1) in Fig. 1
substituting one virtual pion by an $\omega$ meson. The coupling of the
$\omega$ to three pions can be found in \cite{33} and is of the type

\begin{equation} 
V_{\omega \pi \pi \pi} \propto \epsilon_{\mu \nu \alpha \beta} \epsilon^{\mu}
p_1^{\nu} p_2^{\alpha} p_3^{\beta} 
\end{equation}
where $p_i$ are the momenta of the pions and $\epsilon^{\mu}$ the
polarisation vector of the $\omega$. We can expect, in principle,
contributions similar to the $\rho$ meson case. However, at threshold
this term will identically vanish since only the zero
components in Eq. (34) will contribute and they are contracted by an
antisymmetric tensor. Hence we should expect contributions from this term
comparatively smaller than the one from $\rho$ meson exchange discussed above 
and negligible for the total amplitude.  

\section{Results and discussion}

In the first place let us look at the reaction $p p  \rightarrow
p p \pi^+ \pi^-$. We show the cross section as a function of the energy
in Fig. 3. We have separated the contribution of several blocks of
diagrams in the figure. They are calculated using set I of parameters
$c^*_1$, $c^*_2$, but the total contribution is given for both sets I and II
( solid lines ) in order to give an idea of the theoretical uncertainties.
Although the sum of the terms is
done coherently, there is in fact little interference in the total
cross section. The short-dashed curve corresponds to chiral terms,
diagrams (1)-(3) of Fig. 1. As we can see, this contribution is negligible
in this channel. The dash-dotted curve corresponds to the diagrams (9)-(15)
involving only $\Delta$ excitations. We see that this contribution is much
larger than the former one. At low energies it gives a negligible
contribution to the cross section but it rises steeply as a function of the
energy and becomes dominant at large energies. Among all these terms,
$\Delta \Delta$ excitation mechanism of diagram (12) is the largest
above $T_p=1 \ GeV$. The long-dashed curve stands for diagram (8) exciting
$N^*$ and $\Delta$ consecutively. We can see that this term is more relevant
than  the set of $\Delta$ terms at low energies. Finally we show in the
long-short dashed line the contribution of the set of diagrams involving one
$N^*$ excitation followed by a two-pion decay in s-wave, diagrams (4)-(7).
We can see in the figure that this gives by far the largest contribution at low
energies. The sum of all contributions is given by the solid lines.
These lines, corresponding to the two acceptable sets of $c^*_1$, $c^*_2$
parameters, differ by about a factor two at low energies and about $30\, \%$
at $T_p \sim 1 \, GeV$. This sets the level of the theoretical uncertainties
in this reaction. Another reading of these results is that the $p p
\rightarrow  p p \pi^+ \pi^-$ reaction is a more sensitive tool to the
$N^* \rightarrow N (\pi\pi)^{T=0}_{S-wave}$ couplings than the $\pi N
\rightarrow \pi \pi N$ reaction and could be used to put stronger constrains
on them. In the next section we shall see how these
results can be improved. In any case these results have to be seen
with the perspective that by omitting the $N^*$ terms the disagreement
at energies below $T_p = 900 \, MeV$ is larger than two
orders of magnitude. The $\Delta \Delta$ mechanism has been used in
connection with the ABC effect in the
$n p \rightarrow d  + X$ reaction \cite{19}, but at energies
corresponding to $T_p > 1200 \, MeV$ in the elementary reaction.
The strength of this term at these energies is sizable
but we get two other sources of contribution from $N^*$ excitation, as we
have shown, which have about the same strength as the $\Delta \Delta$ term. We
should mention, however, that the consideration of short range
correlations has decreased the contribution of the $\Delta \Delta$  term
with respect to the $\pi$ exchange alone by about a factor three. 

In order to show the relevance of the findings of the isoscalar excitation
of the Roper we show in Fig. 4 the contribution of the $N^*$ 
terms, diagrams (4) and (5) and show there three curves. 
One with the contribution of the term
if we assume a correlated $\pi + \rho$ exchange in the T=1 channel
($V'_L, V'_T$ terms), short-dashed curve,
another one with the contribution assuming only ``$\sigma$'' exchange
, long-dashed curve, and the third one the results with the sum of the two 
exchanges, solid line. As we can see, the contribution with ``$\sigma$'' 
exchange is about one order of magnitude bigger than the one obtained with the 
correlated $\pi + \rho$ exchange. This shows the importance of the novel 
findings on the isoscalar Roper excitation in order to understand the $2 \pi$
production in the $p p $ reaction.
This also gives hopes that the permanent problems in the ABC effect, tied
to the poor angular dependence provided by the $\Delta \Delta$ model
\cite{19,20} could find a solution in the light of this
new interpretation of the $p p \rightarrow p p \pi^+ \pi^-$ reaction.

In Ref. \cite{21} a different approach was followed based on the one
pion exchange model and two mechanisms: one with two pions produced
from the same baryon line and another one with a pion produced in
each baryon line. The ingredients needed there, the
$\pi N \rightarrow \pi \pi N$ and $\pi N \rightarrow \pi N$ amplitudes, were
taken from experimental cross sections, making several
assumptions on how to extrapolate them off shell and summing
incoherently the contribution of the two mechanisms. The model was used at
higher energies that those explored here. Even if phenomenologically
one would be considering in \cite{21}
the terms discussed here with explicit models, it
would only account for the $\pi$ exchange in the terms with
$N^*$ excitation followed by $N \rightarrow N (\pi \pi)^{T=0}_{S-wave}$ or $N^*
\rightarrow \Delta \pi$, while we have shown that the ``$\sigma$'' 
exchange is the dominant piece in the $NN \rightarrow NN^*$ transition.
Also, we have seen that the indirect effect of the short range 
repulsive $NN$ force weakens the $\pi$ exchange contribution in the $2 \pi$
production process. Thus, the experience gathered through the years on
the $NN$ interaction and the pion nucleon and nuclear interaction, 
together with 
the recent findings on isoscalar Roper excitation, have made it
possible the detailed model of the present work, clarifying and
improving the ideas contained in Ref. \cite{21}.

In Fig. 5 we show the results for the $p n \rightarrow p n \pi^+
\pi^-$ reaction with the same meaning as in Fig. 3 and similar results,
although with a larger discrepancy than in the previous case.

In Fig. 6 we show the results off the $p p  \rightarrow p n \pi^+
\pi^0$ channel. This reaction is interesting because the
$N^*$ excitation with $N^* \rightarrow N (\pi \pi)^{T=0}_{S-wave}$
decay shown in diagrams (4) and (5), which were dominant in the 
$p p \rightarrow p p \pi^+ \pi^-$ reaction, do not exist now. 
Diagrams (6) and (7) still
contribute, but they are very small because they involve one $N^* N \pi$
p-wave coupling, which vanishes at threshold and also $\sigma$ exchange is
not allowed. Indeed the $\pi^+ \pi^0$ system can only be in $T = 1,2$ but not 
in $T = 0$. Hence, the mechanism that was dominant in the $p p \rightarrow
p p \pi^+ \pi^-$ reaction at low energies is not present here. In
spite of that, the agreement with the data  is of the same quality
as the one found before for the $p p \rightarrow p p \pi^+ \pi^-$
reaction. Now the dominant terms are those exciting $\Delta$'s. 

In Fig. 7 we show results for the $p n \rightarrow p p \pi^- \pi^0$
reaction. The features are qualitatively similar to those in the previous
channel but the discrepancies are considerably bigger. We note that the
strength of diagram (8) is now comparatively bigger with respect to $\Delta$
excitation terms than in the previous case.

In Fig. 8 the results for the $p p \rightarrow n n \pi^+ \pi^+$ reaction are
shown. Here, at high energies, $\Delta$ terms are still dominant, but below
$1 \ GeV$ chiral terms dominate the amplitudes. This is the only case where
these terms are relatively important. This also means that we should accept
large uncertainties at low energies since, as we saw, there are other terms 
mediated by $\rho$ meson exchange which are of the same order of magnitude 
and were neglected here.

In Fig. 9 we show cross sections for the $p p \rightarrow n n \pi^0 \pi^0$
channel. This is again a channel where the diagrams (4) and (5) are dominant
at low energies, like in the $p p \rightarrow p p \pi^+ \pi^-$ case. Chiral
terms are not drawn since they are below the scale of the figure. In this
case we overestimate the experimental results by about a factor 2-3 although
the quality of the data is not as good as in former cases. 

With these results we exhaust the experimental data and the isospin
independent channels. Indeed, using the isospin analysis of Ref. \cite{14n}
one can see that there are only six independent cross sections. For instance
one can deduce an interesting relationship with the cross sections which is
the following 

\begin{equation}
\begin{array}{c}
2 \sigma(p p \pi^+ \pi^-) - \sigma(p n \pi^+ \pi^0) - 4 \sigma(p p \pi^0
\pi^0) + 2 \sigma(n n \pi^+ \pi^+) \\[2ex]
+ 2 \sigma(p n \pi^+ \pi^-) -2 \sigma(p p
\pi^- \pi^0) - 4 \sigma(p n \pi^0 \pi^0) = 0.
\end{array}
\end{equation}

\noindent
The $N N \pi \pi$ labels stand for the outgoing particles in the given channel.
Incoming ones are fixed by charge conservation. We have also calculated the
$p n \rightarrow p n \pi^0 \pi^0$ cross section with our model, as shown in
Fig. 10, for which there are no experimental data available. We have checked
Eq. (35) independently for the different mechanisms of the model as a test of
consistency which has been passed successfully.   

The model presented here contains only tree level diagrams. Unitarity is not
strictly fulfilled. Actually, imposing unitarity with four particles in the 
final state is less than trivial, as evidenced by the enormous difficulties in 
the case of three body final state \cite{27}. However, one should bare in mind
that as far as we have dominance of a resonant term, the important aspects of
unitarity are included if the proper resonance width is used in the
resonance propagator, as we do. Partial unitarization is accomplished by the
introduction of loops, as done for instance in the $\pi N \rightarrow
\pi \pi N$ reaction for the chiral terms in Ref. \cite{3}. However, we saw
that this sector plays a minor role in the present reaction in view of the
dominant contribution of resonant terms. Unitarization schemes, as the one
of Olsson \cite{28}, have proved to be successful in the two body final
state when a resonant term is dominant and there is a small background. One
multiplies the resonant term by a phase $e^{i \phi}$, with $\phi$ small in
principle, and demand that the resulting amplitude satisfies Watson's theorem,
with the global phase of the amplitude in a particular channel equal
to the one of the final state. The angles $\phi$ needed in the problem of
$\gamma N \rightarrow \pi N $ are of the order of 10$^\circ$ \cite{29}. In
order to have a feeling for what could be the effects of imposing unitarity
in our model, we took the $p p \rightarrow p p \pi^+ \pi^-$ channel and
multiplied the dominant $N^*$ term by $e^{i \phi}$. We see that for values
of $\phi$ up to $20^\circ$ the cross section changes at the level of 1$\%$.
Rough as the procedure appears to be, it gives hints that unitarity is not a
thing to worry much about in the energy domain studied here.

\section{Final state interaction}

We are going to make a qualitative study of the effect of final
state interaction (FSI). Since the energy of the
incoming particles is large at $T_p \geq 800\, MeV$, we take plane waves
for the initial state and look at modifications only from the
interaction of the final particles. Since the low energy region
in the $ p p \rightarrow p p \pi^+ \pi^-$ reaction is dominated by the
$N^* (\pi\pi)^{T=0}_{S-wave}$  contribution, we concentrate on this mechanism 
alone in order to assert the effect of the FSI. For this purpose we substitute

\begin{equation}
\tilde{f} (\vec{q}) = F^2 (q) \frac{1}{q^0\,^2 - \vec{q}\,^2 - m^2_\sigma}
\end{equation}

\noindent
by

\begin{equation}
\int d^3 r \varphi (\vec{r}) \, e^{ \left( i \, \frac{\vec{p}_1
- \vec{p}_2 + \vec{p}_5 + \vec{p}_6}{2}
\, \vec{r}\right)} f (\vec{r})
\end{equation}

\noindent
where $f (\vec{r})$ is the Fourier transform of $\tilde{f} (q)$,

\begin{equation}
f(r) = \int \frac{d^3 q }{(2 \pi)^3} e^{-i  \vec{q} \vec{r}} 
\tilde{f} (\vec{q}\,).
\end{equation}

\noindent
The momenta are chosen according to Fig. 11. By taking a monopole form factor

\begin{equation}
F (q) = \frac{\Lambda^2 - m_\sigma^2}{\Lambda^2 - q^{02} + \vec{q}\,^2}
\end{equation}

\noindent
we find

\begin{equation}
\begin{array}{l}
f (\vec{r}\,) = \frac{\displaystyle{1}}{\displaystyle{4\pi}} \left\{
\frac{\displaystyle{\Lambda^2 - m_\sigma^2}}{\displaystyle{
2 \tilde{\Lambda}}}\,  e^{- \tilde{\Lambda} r}\, 
+ \frac{\displaystyle{e^{- \tilde{\Lambda} r}}}{
\displaystyle{r}} - \frac{\displaystyle{e^{- \tilde{m}_\sigma r}}}{
\displaystyle{r}}
\right\}\\[2ex]
\tilde{\Lambda}\,^2 = \Lambda^2 - (q^0)^2 \quad ; \quad \tilde{m}_\sigma^2
= m_\sigma^2 - (q^0)^2
\end{array}
\end{equation}

\noindent
On the other hand $\varphi (\vec{r})$ is the $ p p$ final wave function,
which for low energies can be written as

\begin{equation}
\varphi(\vec{r}) = e^{\left( i\, \vec{k} \vec{r} 
\right)} + \tilde{\jmath}_0(k, r) - j_0(k r) 
\end{equation}

\noindent
where $\vec{k} = (\vec{p}_4 - \vec{p}_3)/ 2 $, $k = |\vec{k}|$ and
$\tilde{\jmath}_0 (k,r)$ is the interacting $pp$ relative radial
wave function with the boundary condition at $r \rightarrow \infty $

\begin{equation}
\tilde{\jmath}_0 (k,r) \rightarrow e^{i \delta_0} \frac{1}{k r} sin (k r +
\delta_0) 
\end{equation}

\noindent
which we calculate with the Paris potential \cite{22}. Thus we finally
substitute

\begin{equation}
\begin{array}{l}
F^2 (q) \frac{1}{q^{02} - \vec{q}\,^2 - m_\sigma^2} \rightarrow
F^2 (q) \frac{1}{q^{02} - \vec{q}\,^2 - m_\sigma^2}\\[2ex]
+ 4 \pi \int_0^\infty r^2 \, dr \, j_0 (Qr) [\tilde{\jmath}_0 (k, r)
- j_0 (k r))] f(r)
\end{array}
\end{equation}

\noindent
and close to threshold

\begin{equation}
\begin{array}{l}
\vec{Q} \simeq \frac{\vec{p}\,_1 - \vec{p}_2}{2} = 
\vec{p}_1 \quad (CM)\\[2ex]
q^0 \simeq E (p_1) - M \simeq m_\pi
\end{array}
\end{equation}

\noindent
which simplifies much the computations.

The effect of FSI is an increase of the cross section at low energies with
respect to impulse approximation (IA), using plane waves. At high energies,
the approximations of Eqs. (43), (44) are not good but one should expect that 
the IA becomes progressively more accurate. However, one must also take into 
account that some of the effective couplings are already chosen in a way that 
they incorporate FSI effects. This is certainly the case in the correlated 
$\pi + \rho$ exchange which we have discussed. It is also the case in the $N N
\rightarrow N N^*$ transition dominated by the effective ``$\sigma$'' exchange,
since the empirical coupling was obtained in Ref. \cite{11} without
explicit inclusion of FSI. Distortions of proton and pion waves in the
$(\alpha, \alpha')$ reaction was considered in an eikonal approximation in
order to eliminate the $\alpha$ breaking channels. Since the analysis of Ref.
\cite{11} was done at an equivalent $T_p \simeq 1 \ GeV$, we find an
approximate way to account for FSI in the $N^*$ dominated channels in our 
present calculation by multiplying the cross sections by the factor

\begin{equation}
\frac{\sigma^{FSI} (T_p)}{\sigma^{IA} (T_p)} \cdot \frac{\sigma^{IA}
(T_p = 1 GeV)}{\sigma^{FSI} (T_p = 1 GeV)}
\end{equation}

\noindent
up to $T_p = 1 \ GeV$ and we do not modify them at energies higher than 
$T_p = 1 \ GeV$. With these considerations we find the results shown in 
Fig. 12 for the $p p \rightarrow p p \pi^+ \pi^-$ reaction , which should be 
taken as indicative of the role played by FSI. We see that the slope of the 
data is better reproduced when FSI effects are included.  

\section{Conclusions}

We have constructed a model for the $NN \rightarrow \pi \pi NN$ reaction
consisting of the terms appearing from chiral Lagrangians involving
nucleons and pions, plus terms involving the excitation of $\Delta$
 and $N^* $ (1440). In the channels where the two pions can be in a
$T = 0$ state, as $\pi^+ \pi^-$ and $\pi^0 \pi^0$, we find a dominance
of the $N^*$ excitation in one nucleon decaying into $N$ and
$\pi \pi$ in $T = 0$, S-wave. The recent experimental findings about
isoscalar $N^*$ excitation in the $(\alpha, \alpha')$ reaction
on proton targets are used here and one finds that in the
$p p \rightarrow p p \pi^+ \pi^-$, $pn \rightarrow p n \pi^+
\pi^-$ reactions the $NN \rightarrow N N ^*$ transition, driven by
the isoscalar ``$\sigma$'' exchange, and followed by the $N^*
\rightarrow N (\pi \pi)^{T=0}_{S-wave}$ decay largely dominates 
the cross section
at low energies. This is an important finding of the present work which could 
not have been asserted prior to the experimental  observation and 
analysis of the Roper  excitation in the $(\alpha, \alpha')$
reaction. As the energy increases, the $N^*$ excitation followed by
$\Delta \pi$ decay takes also a share of the cross section and so does
the excitation of a $\Delta$ in each of the nucleons, which becomes
dominant at energies $T_p > 1300 \, MeV$. Other terms which are
calculated are found to play a minor role.

A different case is the one of the $p p \rightarrow p n \pi^+ \pi^0$ channel
where the $N^*$ excitation followed by $N^* \rightarrow 
N (\pi \pi)^{T=0}_{S-wave}$ is forbidden. In this case the successive
excitation of two $\Delta$ on the same nucleon, the $\Delta$
excitation on each nucleon, the $N^*$ excitation followed by $\Delta \pi$
decay and even the chiral terms (at low energies) share the strength of
the reaction and one obtains a qualitative agreement with
experiment. However, these same ingredients used in the $p p \rightarrow
p p \pi^+ \pi^-$ reaction but  omitting the
$N^* \rightarrow  N (\pi \pi)^{T=0}_{S-wave}$
decay would give cross sections roughly two orders of
magnitude smaller than experiment at low energies. This gives us a
qualitative idea of the important role played by this mechanism in
this reaction. 

These new mechanisms for the $NN \rightarrow NN \pi \pi$ reaction
are bound to have repercussions in other reactions. The
$ p p \rightarrow p p \pi^0 \pi
^0$ amplitude with one of the two
$\pi^0$ produced in one nucleon and absorbed in the
other one, gives rise to a box diagram that could be relevant for the
$p p \rightarrow p p \pi^0$ reaction close to threshold. Similarly,
the  isotropic piece in $p p \rightarrow p p \pi^+ \pi^-$ coming from
the $N^*$ excitation followed by the $2 \pi$ decay might be the clue to
a better understanding of the ABC effect, which demands such a highly isotropic
amplitude in order to interpret the angular dependence \cite{20}. Steps in
this direction should be encouraged.

\section*{Acknowledgements}

We would like to thank useful discussions with M. J. Vicente Vacas. This
work is partially supported by CICYT contract number AEN 96-1719.
One of us L.A.R. wishes to thank financial support from the Generalitat 
Valenciana.

\newpage
\appendix
\section*{Appendix}
{\vskip.3cm}
\noindent
AMPLITUDES FOR THE $p p \rightarrow p p \pi^{+} \pi^{-}$ CHANNEL.
{\vskip.3cm}
In this channel, the total amplitude can be expressed as 

\begin{equation}
{\cal M}^{(T)}= 
{\cal M}_{{r_3} {r_4} r_1 r_2}({p_3}, {p_4}, p_1, p_2)
-{\cal M}(1 \leftrightarrow 2)
+{\cal M}(3 \leftrightarrow 4, 1 \leftrightarrow 2)
-{\cal M}(3 \leftrightarrow 4)
\end{equation}

\noindent
where the first term in the sum is given below for all mechanisms included
in the calculation. The subindex stands for the number of the diagrams in 
Fig. 1.

\begin{eqnarray}
{\cal M}_{1+2} & = & 
i \frac{1}{6f^{2}} \left(\frac{f_{\pi N N}}{\mu}\right)^{2}
\left\{ \frac{(\vec{\sigma} \cdot \vec{q_1})_{{r_3} r_1}
(\vec{\sigma} \cdot (3\vec{q_1}+\vec{q_2}))_{{r_4} r_2}}
{q_1^2-\mu^2} \right.  \nonumber\\[.4cm]
& & 
\left. -\frac{(\vec{\sigma} \cdot \vec{q}_1)_{{r_3} r_1}}{q_1^2-\mu^2} 
\frac{(\vec{\sigma} \cdot \vec{q}_2)_{{r_4} r_2}}{q_2^2-\mu^2} 
[2(q_1q_2)+ 4(p_5p_6)+ 3\mu^2] \right\} F_{\pi}(q_1) F_{\pi}(q_2)
\\[1cm]
{\cal M}_{3} & = &
-i (4\pi)^2 \frac{\left\{ 2 \frac{\lambda_1}{\mu}-
\frac{\lambda_2}{\mu^{2}}({q_1}^{0}+2{p_5}^{0})\right\}
\delta_{{r_3} r_1} \left\{ 2 \frac{\lambda_1}{\mu}+
\frac{\lambda_2}{\mu^{2}}({q_2}^{0}+2{p_5}^{0})\right\}
\delta_{{r_4} r_2}}{(q_1+p_5)^2 - \mu^2}
\\[1cm]
{\cal M}_{4+5} & = &
-i 2 \frac{\mu^2}{f^{2}} \left(c^*_1 + \frac{{p_5}^0 {p_6}^0}{\mu^2}
c^*_2 \right) \left\{ \frac{f_{\pi N N}}{\mu} \frac{\tilde{f}}{\mu}\left[
\frac{(\vec{\sigma} \cdot \vec{q_2})_{{r_3} r_1} 
(\vec{\sigma} \cdot \vec{q_2})_{{r_4} r_2}}{{\vec{q_2}}^2} \left(V'_L(q_2)
-V'_T(q_2) \right) \right. \right. \nonumber\\[.4cm]
& &
\left. \left.
+V'_T(q_2)(\vec{\sigma})_{{r_3} r_1} \cdot (\vec{\sigma})_{{r_4} r_2} \right]
+g_{\sigma NN} g_{\sigma NN^*} 
\frac{\delta_{{r_3} r_1} \delta_{{r_4} r_2}}{q_2^2-m_{\sigma}^2} 
F_{\sigma}^2(q_2)\right\} \nonumber\\[.4cm]
& &
\times \left[ D_{N^*}(p_5+p_6+{p_3})+D_{N^*}(p_5+p_6-p_1) \right]
\\[1cm]
{\cal M}_{6+7} & = & 0 \qquad (not \ allowed \ by \ isospin \ symmetry)
\\[1cm]
{\cal M}_{8} & = &
i \frac{1}{9} \frac{f^*}{\mu} \frac{g_{\pi N^* \Delta}}{\mu} \left\{ 
\frac{f_{\pi N N}}{\mu} \frac{\tilde{f}}{\mu} \left[ 
\frac{(\vec{\sigma} \cdot \vec{q_2})_{m r_1}
(\vec{\sigma} \cdot \vec{q_2})_{{r_4} r_2}}{{\vec{q_2}}^2} \left( V'_L(q_2)
-V'_T(q_2) \right) \right. \right. \nonumber\\[.4cm]
& &
\left. \left.
+V'_T(q_2)(\vec{\sigma})_{m r_1} \cdot (\vec{\sigma})_{{r_4} r_2} 
\frac{}{} \right]
+g_{\sigma NN} g_{\sigma NN^*}
\frac{\delta_{m r_1} \delta_{{r_4} r_2}}{q_2^2-m_{\sigma}^2}
F_{\sigma}^2(q_2)\right\} \nonumber\\[.4cm]
& &
\times \left\{ 2(\vec{p}_5 \vec{p}_6)\delta_{{r_3}m} 
\left[ D_{\Delta}(p_5+{p_3})+3 D_{\Delta}(p_6+{p_3}) \right] \right.
-i(\vec{\sigma} \cdot [\vec{p_5} \times \vec{p_6}])_{{r_3} m}
\nonumber\\[.4cm]
& &
\left. 
\times \left[ D_{\Delta}(p_5+{p_3})-3 D_{\Delta}(p_6+{p_3}) \right] \right\}
D_{N^*}(p_5+p_6+{p_3})
\end{eqnarray}

\newpage
\begin{eqnarray}
{\cal M}_{9} & = &
i \left( \frac{f^*}{\mu} \right)^2 \frac{f_{\pi N N}}{\mu} 
\frac{f_{\Delta}}{\mu} D_{\Delta}(p_5+{p_6}+{p_3}) \nonumber\\[.4cm]
& &
\times \left[ \frac{1}{9} \left\{
\frac{(\vec{\sigma} \cdot \vec{q_2})_{{r_4} r_2}}{{\vec{q_2}}^2} 
\left[
5i(\vec{q_2} \cdot [ \vec{p_5} \times \vec{p_6} ]) \delta_{{r_3} r_1}
-(\vec{p_5} \cdot \vec{p_6})(\vec{\sigma} \cdot \vec{q_2})_{{r_3} r_1}
\right. \right. \right. \nonumber\\[.4cm]
& &
\left. \left. \left. \hspace{.7cm}
+4(\vec{p_5} \cdot \vec{q_2})(\vec{\sigma} \cdot \vec{p_6})_{{r_3} r_1}
-(\vec{p_6} \cdot \vec{q_2})(\vec{\sigma} \cdot \vec{p_5})_{{r_3} r_1}
\right] (V'_L(q_2)-V'_T(q_2))
\right. \right.  \nonumber\\[.4cm]
& &
\left. \left. \hspace{.7cm}
+\left[ 
5i \delta_{{r_3} r_1} (\vec{\sigma} \cdot [\vec{p_5} \times \vec{p_6}])
_{{r_4} r_2}
-(\vec{p_5} \cdot \vec{p_6})(\vec{\sigma})_{{r_3} r_1} \cdot
(\vec{\sigma})_{{r_4} r_2}
\right. \right. \right. \nonumber\\[.4cm]
& &
\left. \left. \left. \hspace{.7cm}
+4(\vec{\sigma} \cdot \vec{p_6})_{{r_3} r_1}(\vec{\sigma} \cdot
\vec{p_5})_{{r_4} r_2}
-(\vec{\sigma} \cdot \vec{p_5})_{{r_3} r_1}(\vec{\sigma} \cdot
\vec{p_6})_{{r_4} r_2}
\right] V'_T(q_2)
\frac{}{} \right\} D_{\Delta}(p_5+{p_3})
\right. \nonumber\\[.4cm]
& &
\left. \quad
-\frac{1}{6} \left\{ \frac{}{} \vec{p}_5 \leftrightarrow \vec{p}_6 \right\}
D_{\Delta}(p_6+{p_3})
\frac{}{} \right]
\\[1cm]
{\cal M}_{10} & = &
i \frac{4\pi}{3} \left( \frac{f^*}{\mu} \right)^2 \left[ 
\frac{1}{3} \frac{\left\{ 2 \frac{\lambda_1}{\mu}+
\frac{\lambda_2}{\mu^{2}}({q_2}^{0}+2{p_6}^{0})\right\}\delta_{{r_4} r_2}}
{(q_2+p_6)^2-\mu^2} 
\left\{ 2(\vec{p_5} \cdot (\vec{q_2}+\vec{p_6}))\delta_{{r_3} r_1} 
\right. \right. \nonumber\\[.4cm] 
& &
\left. \left. \hspace{2.7cm}
-i(\vec{\sigma} \cdot \left[\vec{p_5} \times (\vec{q_2}+\vec{p_6}) \right])
_{{r_3} r_1} \right\} D_{\Delta}(p_5+{p_3}) 
\right. \nonumber\\[.4cm]
& &
\left.
\hspace{2.2cm} +\frac{\left\{ 2 \frac{\lambda_1}{\mu}-
\frac{\lambda_2}{\mu^{2}}({q_2}^{0}+2{p_5}^{0})\right\}\delta_{{r_4} r_2}}
{(q_2+p_5)^2-\mu^2}
\left\{ 2(\vec{p_6} \cdot (\vec{q_2}+\vec{p_5}))\delta_{{r_3} r_1}
\right. \right. \nonumber\\[.4cm]
& &
\left. \left. \hspace{2.7cm}
-i(\vec{\sigma} \cdot \left[\vec{p_6} \times (\vec{q_2}+\vec{p_5}) \right])
_{{r_3} r_1} \right\} D_{\Delta}(p_6+{p_3})
\frac{}{}\right]
\\[1cm]
{\cal M}_{11} & = &
i \frac{4\pi}{3} \left( \frac{f^*}{\mu} \right)^2 \left[
\frac{\left\{ 2 \frac{\lambda_1}{\mu}+
\frac{\lambda_2}{\mu^{2}}({q_2}^{0}+2{p_6}^{0})\right\}\delta_{{r_4} r_2}}
{(q_2+p_6)^2-\mu^2}
\left\{ 2(\vec{p_5} \cdot (\vec{q_2}+\vec{p_6}))\delta_{{r_3}  r_1}
\right. \right. \nonumber\\[.4cm]
& &
\left. \left. \hspace{2.7cm}
+i(\vec{\sigma} \cdot \left[\vec{p_5} \times (\vec{q_2}+\vec{p_6}) \right])
_{{r_3} r_1} \right\} D_{\Delta}(p_1-p_5)
\right. \nonumber\\[.4cm]
& &
\left.
\hspace{2.2cm} +\frac{1}{3} \frac{\left\{ 2 \frac{\lambda_1}{\mu}-
\frac{\lambda_2}{\mu^{2}}({q_2}^{0}+2{p_5}^{0})\right\}\delta_{{r_4} r_2}}
{(q_2+p_5)^2-\mu^2}
\left\{ 2(\vec{p_6} \cdot (\vec{q_2}+\vec{p_5}))\delta_{{r_3} r_1}
\right. \right. \nonumber\\[.4cm]
& &
\left. \left. \hspace{2.7cm}
+i(\vec{\sigma} \cdot \left[\vec{p_6} \times (\vec{q_2}+\vec{p_5}) \right])
_{{r_3} r_1} \right\} D_{\Delta}(p_1-p_6)
\frac{}{}\right]
\end{eqnarray}

\newpage
\begin{eqnarray}
{\cal M}_{12} & = &
i \frac{1}{3}\frac{1}{9} \left( \frac{f^*}{\mu} \right)^4 \left(
\frac{1}{(\vec{q_2}+\vec{p_6})^2}
\left\{ 2(\vec{p}_6 \cdot (\vec{q}_2+\vec{p}_6))\delta_{{r_4} r_2}
-i(\vec{\sigma} \cdot \left[\vec{p_6} \times (\vec{q_2}+\vec{p_6})\right])
_{{r_4} r_2} \right\} \right. \nonumber\\[.4cm]
& &
\left.
\times \left\{ 2(\vec{p}_5 \cdot (\vec{q}_2+\vec{p}_6))\delta_{{r_3} r_1}
-i(\vec{\sigma} \cdot \left[\vec{p_5} \times (\vec{q_2}+\vec{p_6})\right])
_{{r_3} r_1} \right\} 
(V'_L(q_2+p_6)-V'_T(q_2+p_6))
\right. \nonumber\\[.4cm]
& &
\left.
+\left\{ (\vec{p}_5 \cdot \vec{p}_6)(-2\delta_{{r_4} r_1}\delta_{{r_3} r_2}
+5\delta_{{r_3} r_1}\delta_{{r_4} r_2})
+(\vec{\sigma} \cdot \vec{p}_5)_{{r_4} r_2}(\vec{\sigma} \cdot
\vec{p}_6)_{{r_3} r_1}
\right. \right. \nonumber\\[.4cm]
& &
\left. \left.
+2i(\vec{\sigma} \cdot \left[\vec{p}_5 \times \vec{p}_6 \right])_{{r_4} r_2} 
\delta_{{r_3} r_1}
-2i(\vec{\sigma} \cdot \left[\vec{p}_5 \times \vec{p}_6 \right])_{{r_3} r_1}
\delta_{{r_4} r_2} \right\} 
V'_T(q_2+p_6) \frac{}{}\right)
\nonumber\\[.4cm]
& &
\times D_{\Delta}({p_4}+p_6)D_{\Delta}({p_3}+p_5)
\\[1cm]
{\cal M}_{13} & = &
i \frac{1}{3}\frac{1}{9} \left( \frac{f^*}{\mu} \right)^4 \left(
\frac{1}{(\vec{q_2}+\vec{p_6})^2}
\left\{ 2(\vec{p}_6 \cdot (\vec{q}_2+\vec{p}_6))\delta_{{r_4} r_2}
+i(\vec{\sigma} \cdot \left[\vec{p_6} \times (\vec{q_2}+\vec{p_6})\right])
_{{r_4} r_2} \right\} \right. \nonumber\\[.4cm]
& &
\left.
\times \left\{ 2(\vec{p}_5 \cdot (\vec{q}_2+\vec{p}_6))\delta_{{r_3} r_1}
+i(\vec{\sigma} \cdot \left[\vec{p_5} \times (\vec{q_2}+\vec{p_6})\right])
_{{r_3} r_1} \right\}
(V'_L(q_2+p_6)-V'_T(q_2+p_6))
\right. \nonumber\\[.4cm]
& &
\left.
+\left\{ (\vec{p}_5 \cdot \vec{p}_6)(-2\delta_{{r_4} r_1}\delta_{{r_3} r_2}
+5\delta_{{r_3} r_1}\delta_{{r_4} r_2})
+(\vec{\sigma} \cdot \vec{p}_5)_{{r_4} r_2}(\vec{\sigma} \cdot
\vec{p}_6)_{{r_3} r_1}
\right. \right. \nonumber\\[.4cm]
& &
\left. \left.
-2i(\vec{\sigma} \cdot \left[\vec{p}_5 \times \vec{p}_6 \right])_{{r_4} r_2}
\delta_{{r_3} r_1}
+2i(\vec{\sigma} \cdot \left[\vec{p}_5 \times \vec{p}_6 \right])_{{r_3} r_1}
\delta_{{r_4} r_2} \right\}
V'_T(q_2+p_6) \frac{}{}\right)
\nonumber\\[.4cm]
& &
\times D_{\Delta}(p_2-p_6)D_{\Delta}(p_1-p_5)
\\[1cm]
{\cal M}_{14} & = &
i \frac{1}{9}\frac{1}{9} \left( \frac{f^*}{\mu} \right)^4 \left(
\frac{1}{(\vec{q_2}+\vec{p_6})^2}
\left\{ 2(\vec{p}_6 \cdot (\vec{q}_2+\vec{p}_6))\delta_{{r_4} r_2}
+i(\vec{\sigma} \cdot \left[\vec{p_6} \times (\vec{q_2}+\vec{p_6})\right])
_{{r_4} r_2} \right\} \right. \nonumber\\[.4cm]
& &
\left.
\times \left\{ 2(\vec{p}_5 \cdot (\vec{q}_2+\vec{p}_6))\delta_{{r_3} r_1}
-i(\vec{\sigma} \cdot \left[\vec{p_5} \times (\vec{q_2}+\vec{p_6})\right])
_{{r_3} r_1} \right\}
(V'_L(q_2+p_6)-V'_T(q_2+p_6))
\right. \nonumber\\[.4cm]
& &
\left.
+\left\{ (\vec{p}_5 \cdot \vec{p}_6)(2\delta_{{r_4} r_1}\delta_{{r_3} r_2}
+3\delta_{{r_3} r_1}\delta_{{r_4} r_2})
-(\vec{\sigma} \cdot \vec{p}_5)_{{r_4} r_2}(\vec{\sigma} \cdot
\vec{p}_6)_{{r_3} r_1}
\right. \right. \nonumber\\[.4cm]
& &
\left. \left.
-2i(\vec{\sigma} \cdot \left[\vec{p}_5 \times \vec{p}_6 \right])_{{r_4} r_2}
\delta_{{r_3} r_1}
-2i(\vec{\sigma} \cdot \left[\vec{p}_5 \times \vec{p}_6 \right])_{{r_3} r_1}
\delta_{{r_4} r_2} \right\}
V'_T(q_2+p_6) \frac{}{}\right)
\nonumber\\[.4cm]
& &
\times D_{\Delta}(p_2-p_6)D_{\Delta}({p_3}+p_5)
\end{eqnarray}

\newpage
\begin{eqnarray}
{\cal M}_{15} & = &
i \frac{1}{9} \left( \frac{f^*}{\mu} \right)^4 \left(
\frac{1}{(\vec{q_2}+\vec{p_6})^2}
\left\{ 2(\vec{p}_6 \cdot (\vec{q}_2+\vec{p}_6))\delta_{{r_4} r_2}
-i(\vec{\sigma} \cdot \left[\vec{p_6} \times (\vec{q_2}+\vec{p_6})\right])
_{{r_4} r_2} \right\} \right. \nonumber\\[.4cm]
& &
\left.
\times \left\{ 2(\vec{p}_5 \cdot (\vec{q}_2+\vec{p}_6))\delta_{{r_3} r_1}
+i(\vec{\sigma} \cdot \left[\vec{p_5} \times (\vec{q_2}+\vec{p_6})\right])
_{{r_3} r_1} \right\}
(V'_L(q_2+p_6)-V'_T(q_2+p_6))
\right. \nonumber\\[.4cm]
& &
\left.
+\left\{ (\vec{p}_5 \cdot \vec{p}_6)(2\delta_{{r_4} r_1}\delta_{{r_3} r_2}
+3\delta_{{r_3} r_1}\delta_{{r_4} r_2})
-(\vec{\sigma} \cdot \vec{p}_5)_{{r_4} r_2}(\vec{\sigma} \cdot
\vec{p}_6)_{{r_3} r_1}
\right. \right. \nonumber\\[.4cm]
& &
\left. \left.
+2i(\vec{\sigma} \cdot \left[\vec{p}_5 \times \vec{p}_6 \right])_{{r_4} r_2}
\delta_{{r_3} r_1}
+2i(\vec{\sigma} \cdot \left[\vec{p}_5 \times \vec{p}_6 \right])_{{r_3} r_1}
\delta_{{r_4} r_2} \right\}
V'_T(q_2+p_6) \frac{}{}\right)
\nonumber\\[.4cm]
& &
\times D_{\Delta}({p_4}+p_6)D_{\Delta}(p_1-p_5)
\end{eqnarray}

\noindent
In these expressions $p_1$, $p_2$ ($p=(p^0, \vec{p})$) are the momenta of the
incoming nucleons while ${p_3}$, ${p_4}$ are the momenta of the outgoing
ones, $p_5, p_6$ are $\pi^-$ and $\pi^+$ momenta respectively and $q_{1(2)}$
denote ${p_{3(4)}}-p_{1(2)}$. Here ${r_3},{r_4}, r_1, r_2, m=(1,2)$ are
spin indices and a sum in $m$ is understood in Eq. (51) $V'_L$ and $V'_T$
are the longitudinal and transverse parts of the $N N$ interaction as defined
in Eq. (22), $F_{\pi}(q)$ and $F_{\sigma}(q)$ are pion and sigma form
factors respectively. Finally $D_{N^*}(p)$ and $D_{\Delta^*}(p)$ stand for
$N^*$ and $\Delta$ propagators, given by the following expressions 

\begin{equation}
D_i(p)=\frac{1}{\sqrt{p^2}- M_i+ \frac{1}{2} i \Gamma_i(p)} 
\frac{M_i}{\sqrt{M_i^2+\vec{p}_i^2}} \quad ; \qquad i=(N^*,\Delta).
\end{equation}

\noindent
Here $\Gamma_i(p)$ denotes the width of the resonance \cite{11}. 
\newpage
\section*{Figure Captions}
\bigskip
\parindent 0 cm

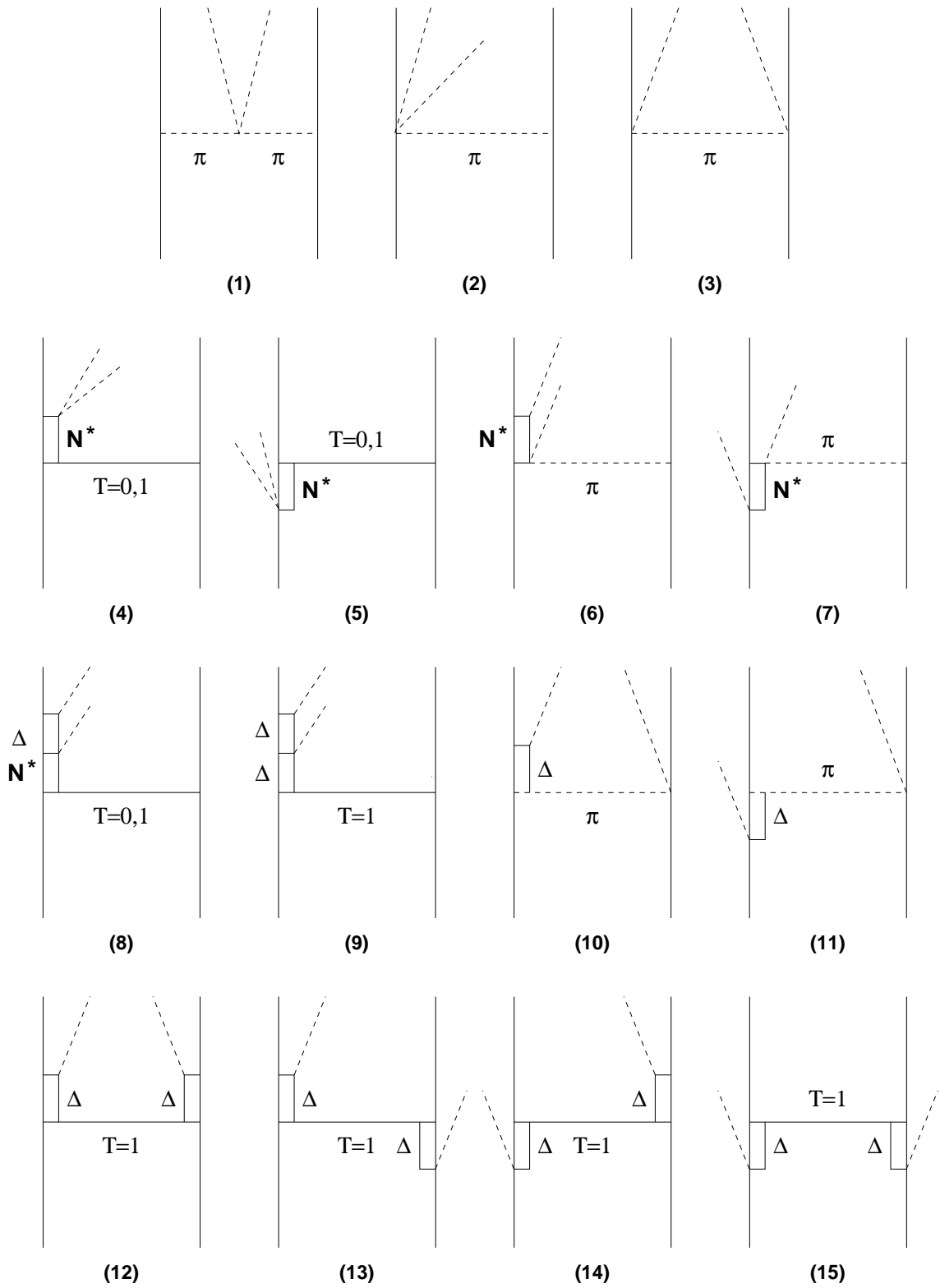
{\bf Fig. 1.} Complete set of Feynman diagrams of our model.
\medskip

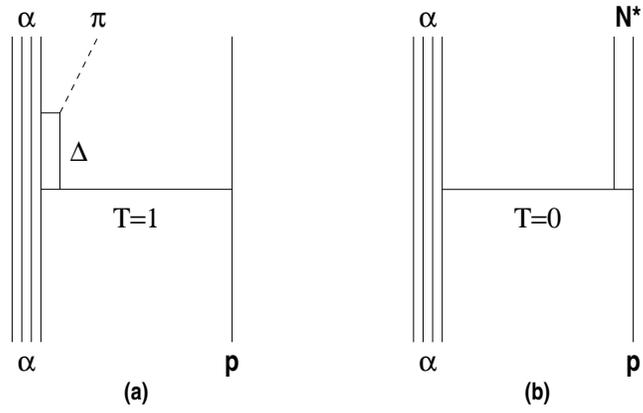
{\bf Fig. 2.} Dominant mechanisms in the analysis of $(\alpha, \alpha')$
performed in Ref. \cite{11}. (a) $\Delta$ excitation in the projectile, (b)
Roper excitation in the target.
\medskip

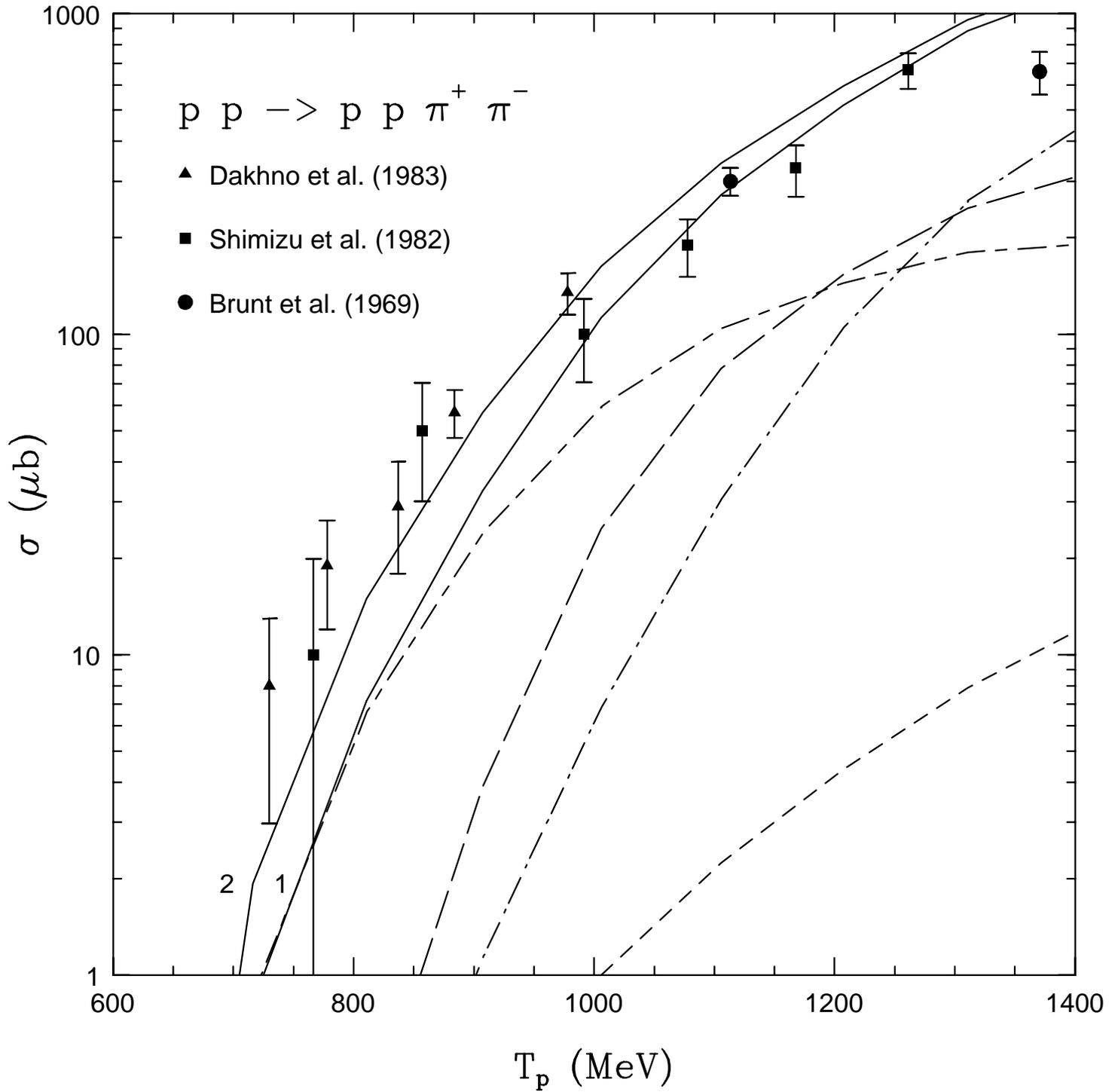
{\bf Fig. 3.} Total cross section for the $p p \rightarrow p p \pi^+ \pi^-$
channel as a function of the incoming proton kinetic energy in the laboratory
frame. Solid line, total ( line labelled 1 for set I and 2 for set II of 
$c^*_1$, $c^*_2$ parameters )
; long-short dashed line, $N^* \rightarrow  N (\pi
\pi)^{T=0}_{S-wave}$; long-dashed line, $N^* \rightarrow \Delta \pi$;
dash-dotted line, $\Delta$ excitation mechanisms; short-dashed line,
non-resonant terms from diagrams (1)-(3). The partial contributions are
calculated with set I. Experimental data are taken from 
Refs. \cite{12n,13n,14n}.
\medskip

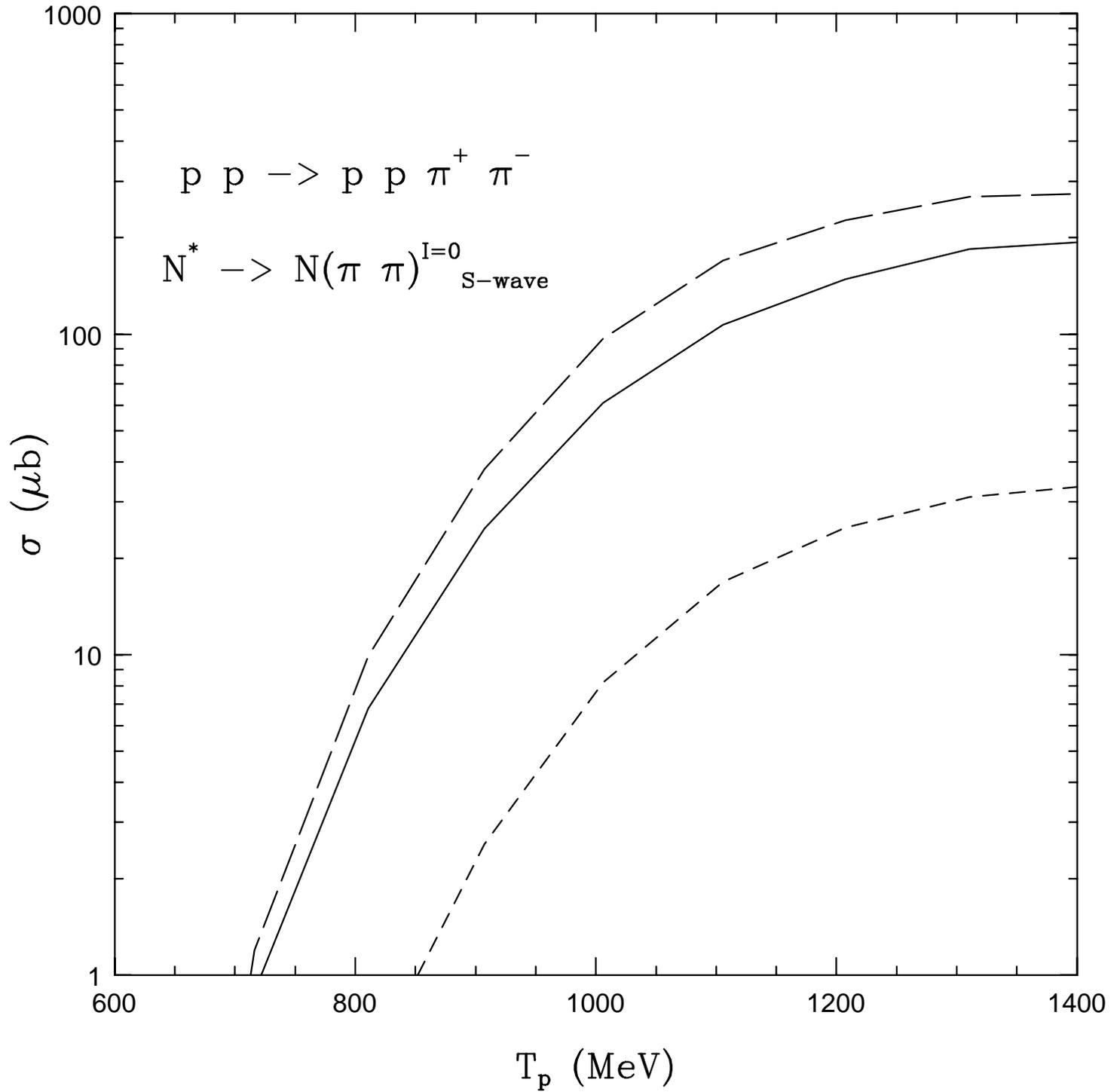
{\bf Fig. 4.} $N^* \rightarrow  N (\pi\pi)^{T=0}_{S-wave}$ contribution to
$p p \rightarrow p p \pi^+ \pi^-$ total cross section ( set I )
separated in $T=0$ ``$\sigma$'' exchange
( long-dashed line ) and $T=1$ correlated $\pi + \rho$ exchange
(short-dashed line ) pieces. The solid line gives their sum.

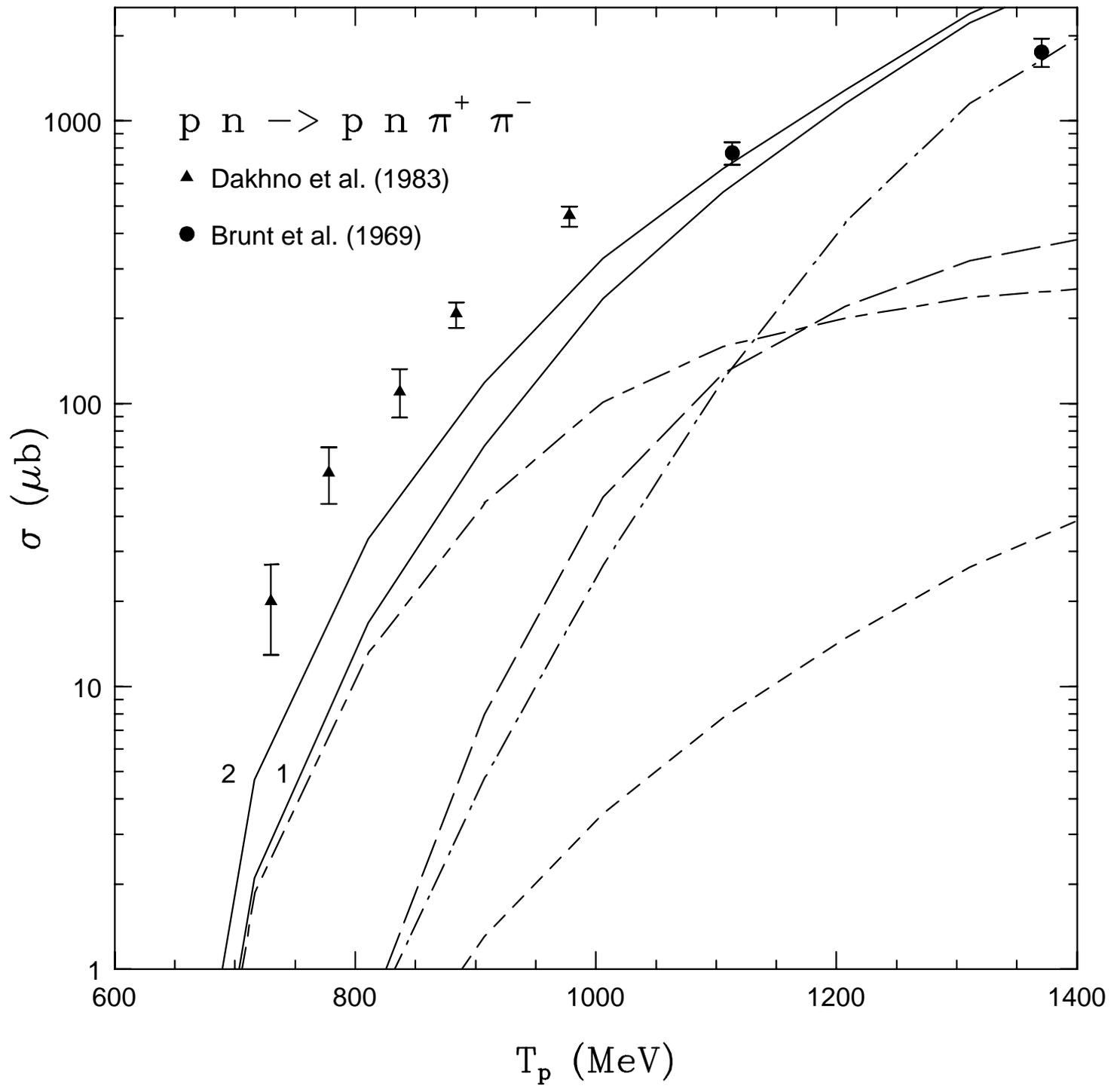
{\bf Fig. 5.} Same as Fig. 3 for $p n \rightarrow p n \pi^+ \pi^-$
\medskip

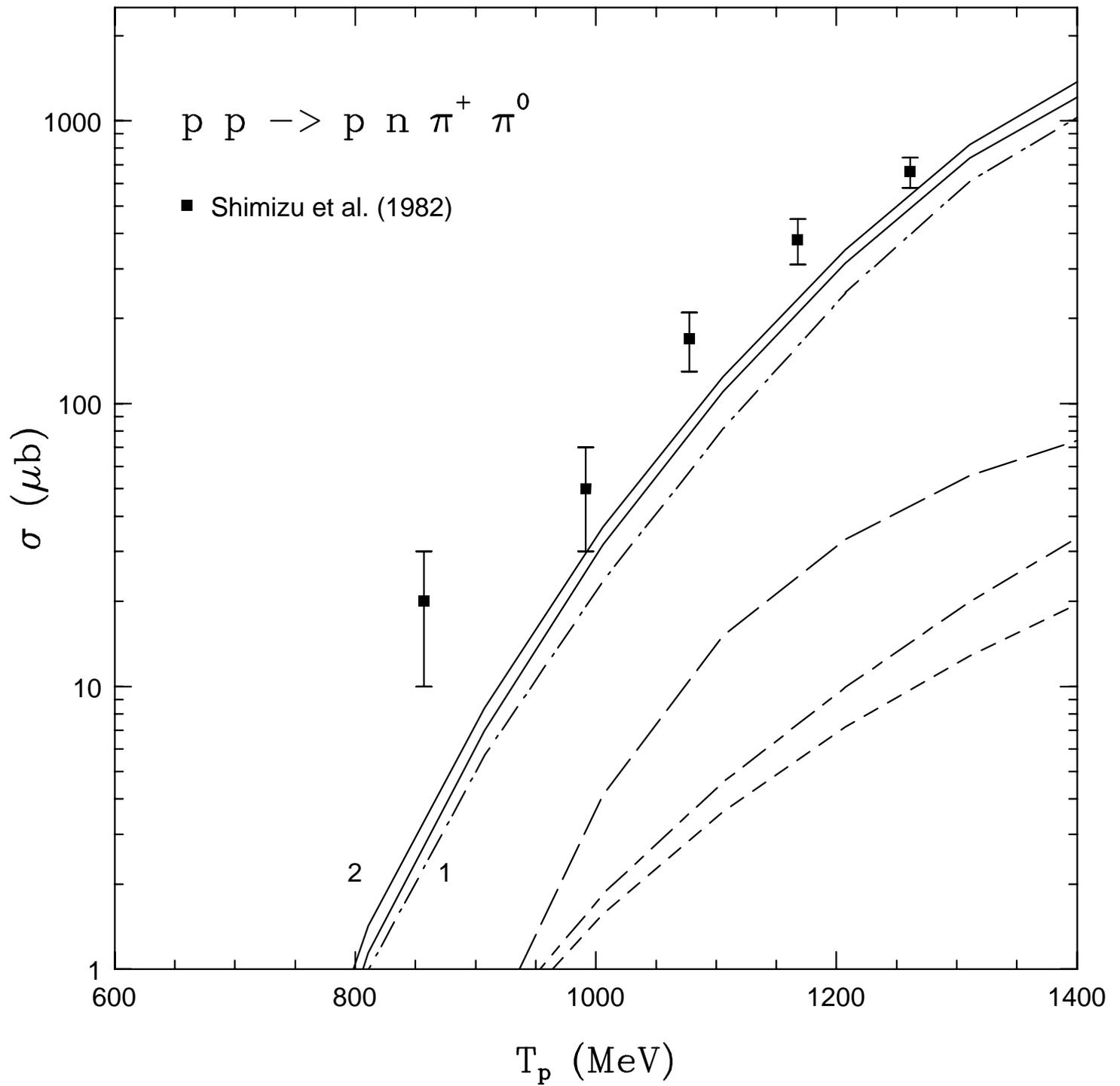
{\bf Fig. 6.} Same as Fig. 3 for $p p \rightarrow p n \pi^+ \pi^0$
\medskip

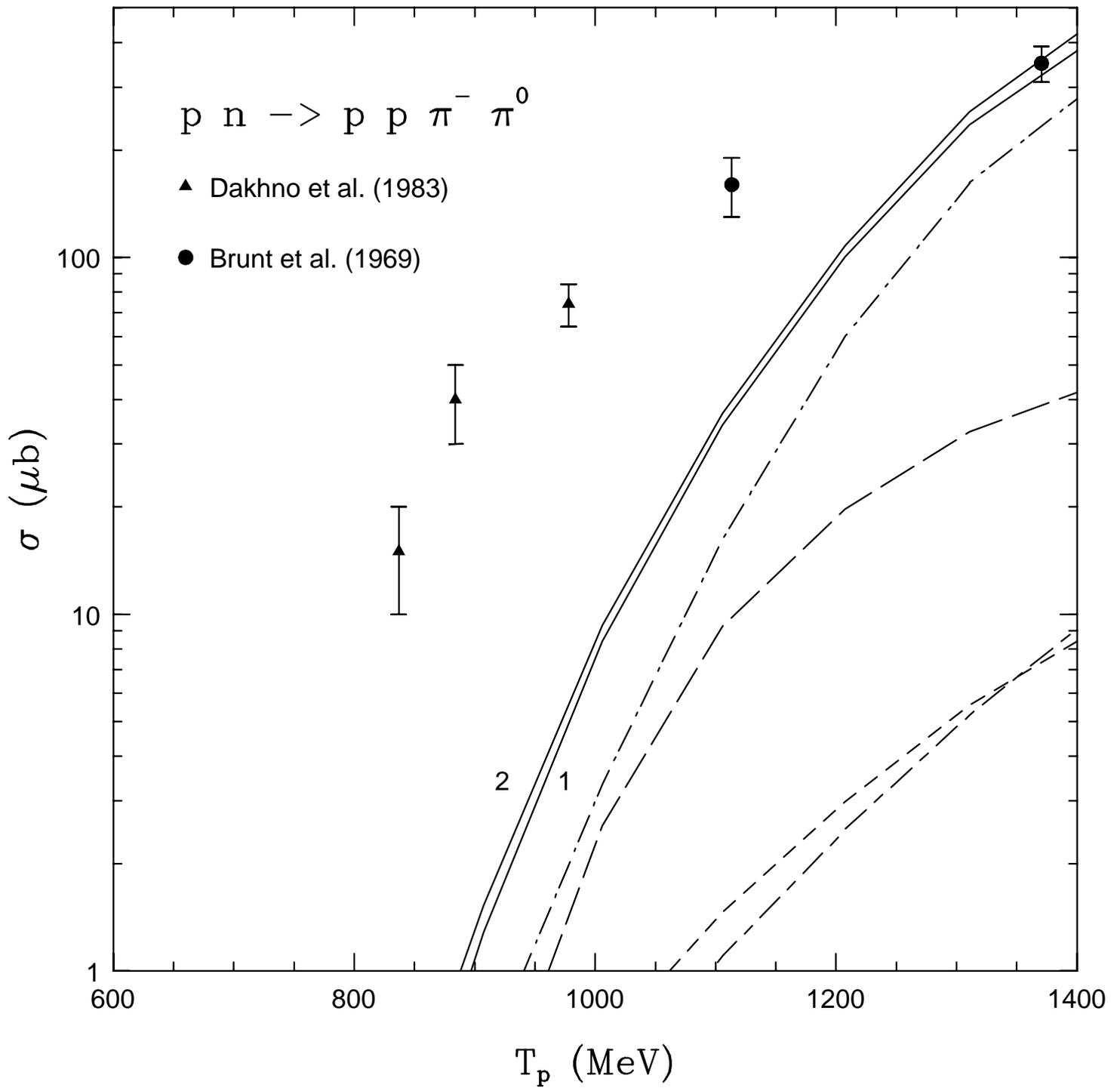
{\bf Fig. 7.} Same as Fig. 3 for $p n \rightarrow p p \pi^- \pi^0$
\medskip

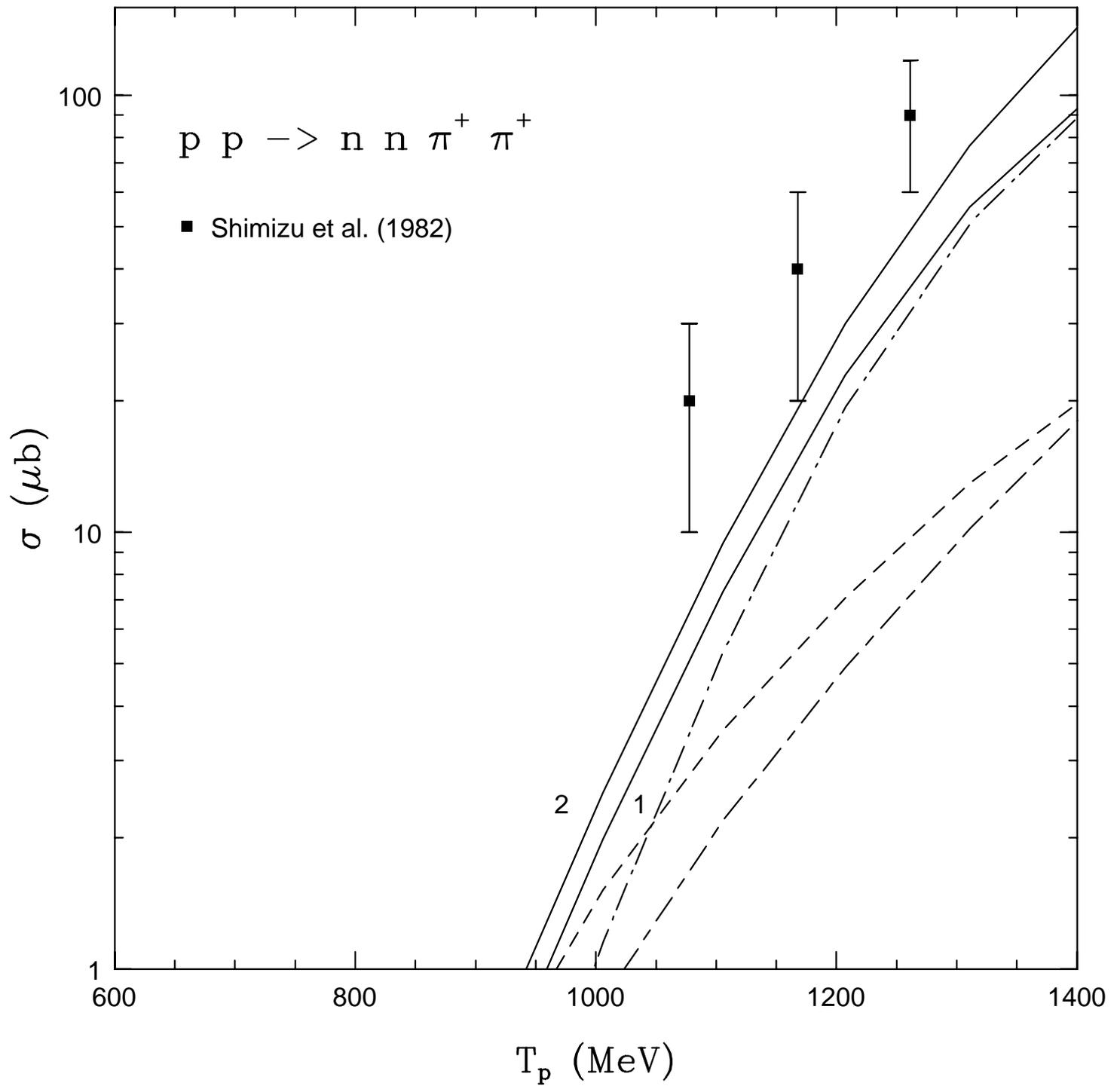
{\bf Fig. 8.} Same as Fig. 3 for $p p \rightarrow n n \pi^+ \pi^+$
\medskip

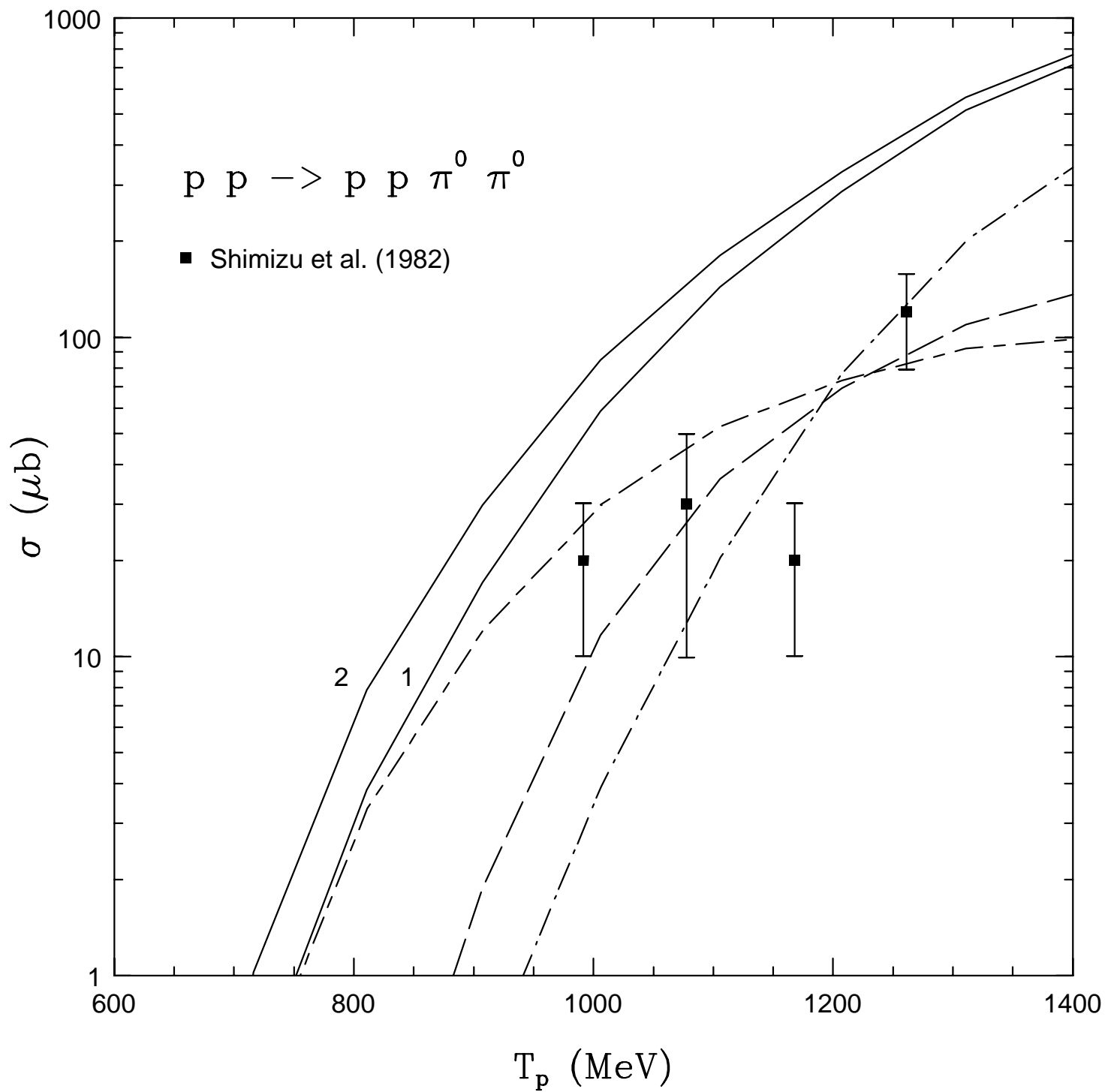
{\bf Fig. 9.} Same as Fig. 3 for $p p \rightarrow p p \pi^0 \pi^0$
\medskip

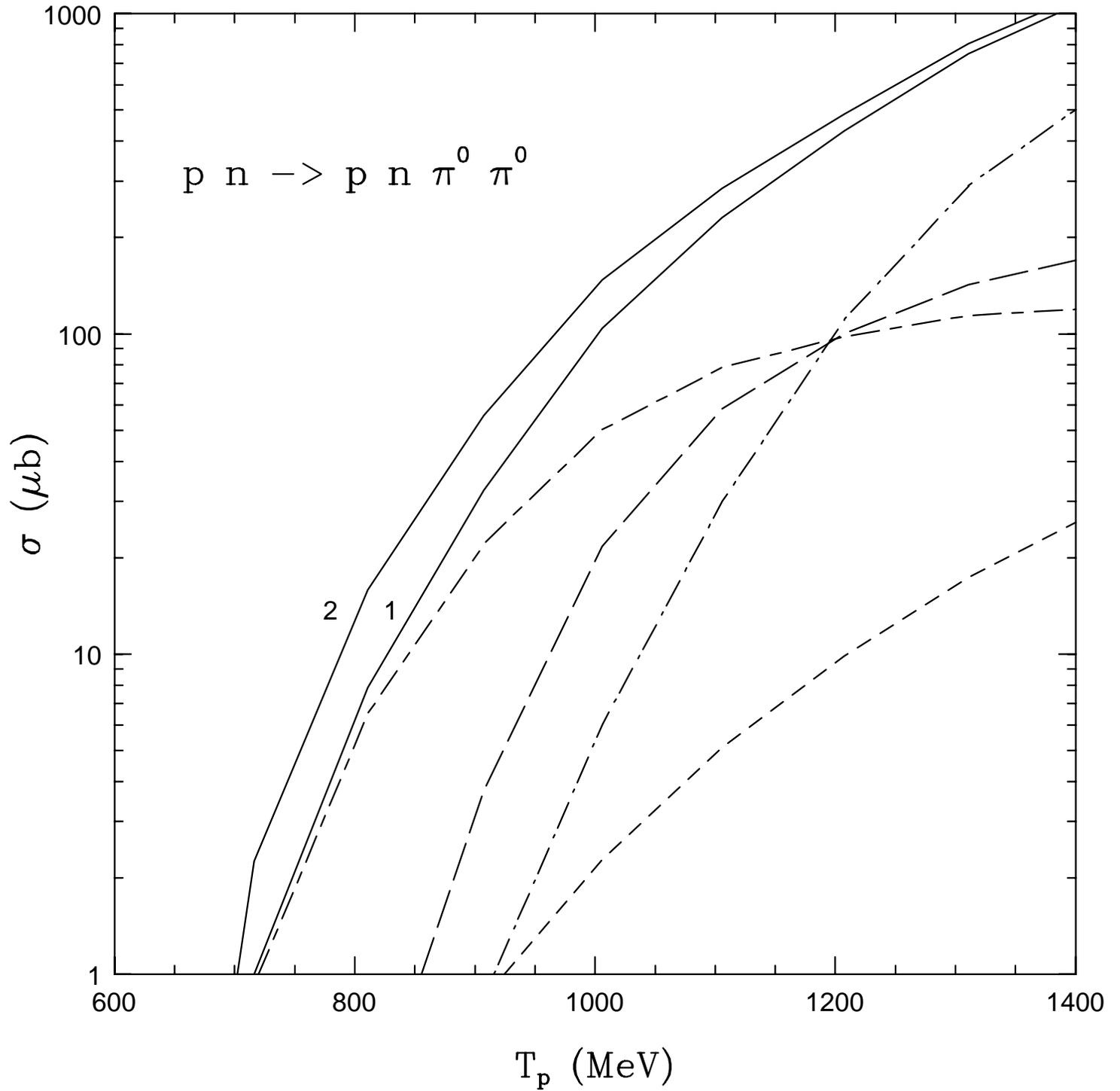
{\bf Fig. 10.} Same as Fig. 3 for $p n \rightarrow p n \pi^0 \pi^0$
\medskip

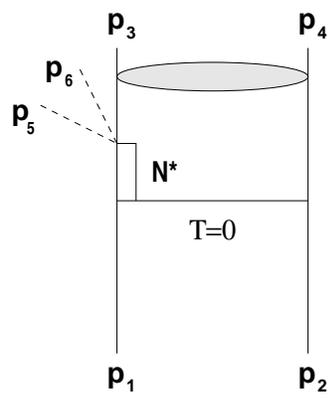
{\bf Fig. 11.} Dominant mechanism in the $p p \rightarrow p p \pi^+ \pi^-$
reaction at low energies with final state interaction included.
\medskip

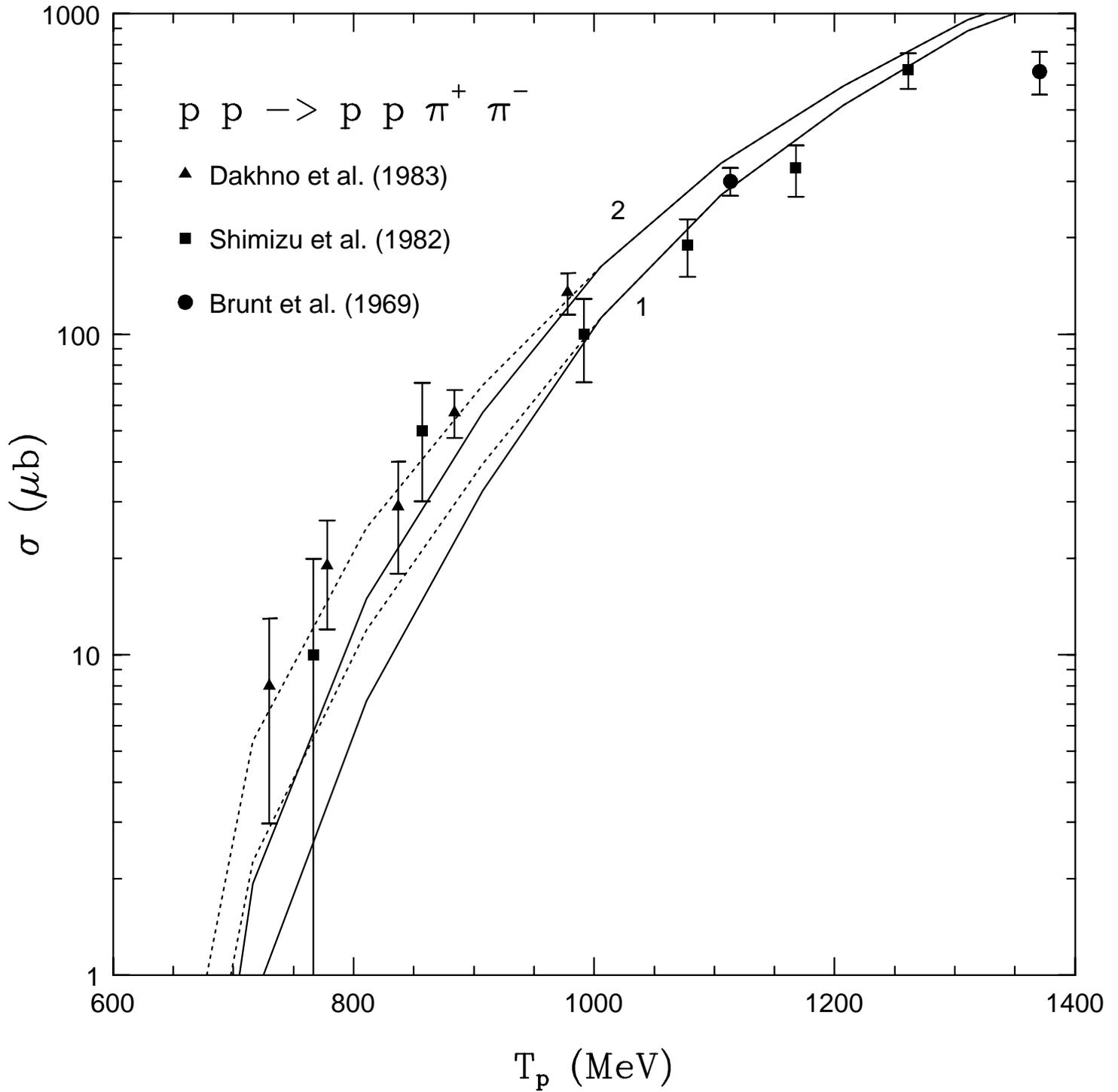
{\bf Fig. 12.} Total cross section for the $p p \rightarrow p p \pi^+
\pi^-$ channel with ( dotted lines ) and without ( solid lines ) final state
interaction for both sets I and II. 

\end{document}